\documentclass[11pt]{article}

\usepackage{amssymb,amsmath,amsfonts}
\usepackage{graphicx,graphics}
\DeclareGraphicsExtensions{.jpg,.jpeg,.pdf,.png,.mps,.eps,.ps,.gif}
\usepackage{color}
\usepackage{framed}
\graphicspath{{matlab/}}
\newcommand{\rem}[1]{}
\DeclareMathAlphabet{\mathbi}{OML}{cmm}{b}{it} 
\newcommand{\non}{\nonumber}

\newtheorem{proposition}{Proposition}

\newcommand{\bx}{\mathbi{x}}

\newcommand{\bel}{\begin{equation}\label}
\newcommand{\ee}{\end{equation}}
\newcommand{\beq}{\begin{eqnarray}\label} 
\newcommand{\eeq}{\end{eqnarray}} 
\newcommand{\bc}{\begin{center}} 
\newcommand{\ec}{\end{center}} 
\newcommand{\ben}{\begin{enumerate}}
\newcommand{\een}{\end{enumerate}}
\newcommand{\bit}{\begin{itemize}}
\newcommand{\eit}{\end{itemize}}
\newcommand{\I}{\int_{\mathcal{V}}}

\newcommand{\bdf}{\mathbi{f}}

\newcommand{\bu}{\mbox{\boldmath$u$}}

\newcommand{\bom}{\mbox{\boldmath$\omega$}}
\newcommand{\bphi}{\mbox{\boldmath$\phi$}}

\newcommand{\bk}{\mbox{\boldmath$k$}}

\newcommand\shalf{\ensuremath{{\scriptstyle\frac{1}{2}}}}

\newcommand\twothirds{\ensuremath{{\scriptstyle\frac{2}{3}}}}

\newcommand{\lamin}{\lambda_{\rm min}}
\makeatletter
\@addtoreset{equation}{section}

\makeatother

\begin{document}
\sf
\bc
\textbf{\Large\color{blue}Regimes of nonlinear depletion and regularity\\
in the 3D Navier-Stokes equations}
\par\vspace{5mm}\noindent
%%%%%%%%%%
\textbf{John D. Gibbon}\footnote{\sf j.d.gibbon@ic.ac.uk~~~http://www2.imperial.ac.uk/$\sim$jdg}
\par\vspace{1mm}\noindent
Department of Mathematics, Imperial College London, London SW7 2AZ, UK
\par\vspace{3mm}\noindent
%%%%%%%
\textbf{Diego A. Donzis}\footnote{\sf donzis@tamu.edu~~~http://people.tamu.edu/$\sim$donzis}
\par\vspace{1mm}\noindent
Department of Aerospace Engineering,\\
Texas A\&M University, College Station, Texas, TX 77840, USA
%%%%%%%
\par\vspace{3mm}\noindent
\textbf{Anupam Gupta}\footnote{\sf anupam1509@gmail.com~~~http://people.roma2.infn.it/$\sim$agupta}
\par\vspace{1mm}\noindent
Department of Physics, University of Rome `Tor Vergata', 00133 Roma, Italy
\par\vspace{3mm}\noindent
%%%%%%%
\textbf{Robert M. Kerr}\footnote{\sf robert.kerr@warwick.ac.uk~~~http://www.eng.warwick.ac.uk/staff/$\sim$rmk}
\par\vspace{1mm}\noindent
Department of Mathematics, University of Warwick, Coventry CV4 7AL, UK
\par\vspace{3mm}\noindent
%%%%%%%
\textbf{Rahul Pandit}\footnote{\sf rahul@physics.iisc.ernet.in~~~http://www.physics.iisc.ernet.in/$\sim$rahul}
\par\vspace{1mm}\noindent
Department of Physics, Indian Institute of Science, Bangalore 560 012, India\\
and\\
Jawaharlal Nehru Centre for Advanced Scientific Research, Bangalore, India
\par\vspace{3mm}\noindent
%%%%%%%
\textbf{Dario Vincenzi}\footnote{\sf dario.vincenzi@unice.edu~~~http://math.unice.fr/$\sim$vincenzi}
\par\vspace{1mm}\noindent
Laboratoire Jean Alexandre Dieudonn\'{e}, Universit\'{e} Nice Sophia Antipolis,\\
CNRS, UMR 7351, 06100 Nice, France
\ec

\thispagestyle{empty}

\par\vspace{5mm}

\begin{abstract}
The periodic $3D$ Navier-Stokes equations are analyzed in terms of dimensionless, scaled, $L^{2m}$-norms 
of vorticity $D_{m}$ ($1 \leq m < \infty$). The first in this hierarchy, $D_{1}$, is the global enstrophy. Three 
regimes naturally occur in the $D_{1}-D_{m}$ plane.  Solutions in the first regime, which lie between two 
concave curves, are shown to be regular, owing to strong nonlinear depletion. Moreover, numerical experiments 
have suggested, so far, that all dynamics lie in this heavily depleted regime \cite{DGGKPV13}\,; new numerical 
evidence for this is presented. Estimates for the dimension of a global attractor and a corresponding inertial range 
are given for this regime. However, two more regimes can theoretically exist. In the second, which lies between 
the upper concave curve and a line, the depletion is insufficient to regularize solutions, so no more than Leray's 
weak solutions exist. In the third, which lies above this line, solutions are regular, but correspond to extreme 
initial conditions. The paper ends with a discussion on the possibility of transition between these regimes. 
\end{abstract}

\newpage
%%%%%%%%%%%%%
\section{\sf Introduction}\label{introduction}

Kolmogorov's phenomenological statistical theory of turbulence, based on a set of axioms, displays certain well-known 
characteristics, such as a $k^{-5/3}$ spectrum in an inertial range with a wavenumber cut-off at $L^{-1}Re^{3/4}$, 
together with a dissipation range beyond this [2-5]. %\cite{Frisch95,SA97,BMV08,PPR09}.   
In contrast, from the perspective of Navier-Stokes analysis, much remains open in the three-dimensional case [6-19]. 
%\cite{vW83,Giga86,KS97,ESS03,CF88,IDDS,DG95,FMRT01,KZ06,KZ07,CT08,CRD09,CS11,BF12}. 
A proof of the existence and uniqueness of solutions is missing so the existence of a global attractor remains an open question [10-13]. %\cite{CF88,IDDS,DG95,FMRT01}. 
Moreover, characteristics of an energy spectrum, such as its steepness and wavenumber cut-off, are hard to extract from a 
time-evolving PDE. An interesting question is whether numerical experiments on the Navier-Stokes equations can inform 
the analysis by suggesting a new and different way of looking at Navier-Stokes turbulence? In the early days of Navier-Stokes 
simulations [20-25] %\cite{OrsPat,Rog81,RMK85,EP88,JWSR93,MM98} 
less resolution was available but, in recent years, several very large simulations (up to a maximum of $4096^{3}$) have 
been performed [26-32]. %\cite{KT05,IGK2009,DYS2008,DY2010,YDS2012,RMK2012a,RMK2013a}. 
The data from two of these, together with additional computations, are used in an attempt to understand the behaviour of 
the solutions from a range of initial conditions. 
\par\smallskip%\noindent
The variables that will be used in this paper are defined in terms of the Navier-Stokes vorticity field $\bom = \mbox{curl}\,
\bu$ in the following manner [1,\,33-36]\,: %\cite{JDGPRS10,JDGCMS11,JDGJMP12,JDGiutam13,DGGKPV13}\,:
\beq{s1}
\Omega_{m}(t) = \left(L^{-3}\I |\bom|^{2m}dV\right)^{1/2m}\,;
\qquad D_{m} = \left(\varpi_{0}^{-1}\Omega_{m}\right)^{\alpha_{m}}\,;
\qquad\alpha_{m} = \frac{2m}{4m-3}\,,
\eeq
where $\varpi_{0} = \nu L^{-2}$ is the frequency on the periodic box $[0,\,L]^{3}$. Note that $D_{1} = \left(\varpi_{0}^{-1}
\Omega_{1}\right)^{2}$ is proportional to the $H_{1}$-norm of the velocity field. 
A recent set of numerical experiments, using a variety of initial conditions, each with periodic boundary conditions \cite{DGGKPV13}, 
has suggested that the $D_{m}$ are ordered on a descending scale such that $D_{m+1} < D_{m}$ for $m\geq 1$. In itself this is not 
surprising\,: while H\"older's inequality necessarily enforces the $\Omega_{m}$ to be ordered on an ascending scale such that 
$\Omega_{m} \leq \Omega_{m+1}$, the decreasing nature of the $\alpha_{m}$ means that if the $\Omega_{m}$ are bunched 
sufficiently close, the ordering of the $D_{m}$ could easily be the reverse of the $\Omega_{m}$, as indeed is observed numerically. 
What is more surprising is the observed strong separation on a logarithmic scale in the descending sequence of the $D_{m}$, in 
particular from $D_{1}$. This separation is observed to be of the form\footnote{\sf The exponent has been changed to $A_{m}$ 
from $a_{m}$ in \cite{DGGKPV13} to avoid confusion with $\alpha_{m}$.} (see \S\ref{threereg})
\bel{dep1a}
D_{1}^{\alpha_{m}/2} \leq D_{m} \leq D_{1}^{A_{m}}\,,\qquad\qquad m\geq 2\,,
\ee
where $\shalf\alpha_{m} < A_{m}(t) < \shalf$\,: the lower bound arises from $\Omega_{1} \leq \Omega_{m}$ expressed in 
the $D_{m}$-notation. 
\par\smallskip%\noindent
The main intention of this paper is to investigate how the numerically observed depletion in (\ref{dep1a}) severely reduces the 
strength of the vortex stretching, thereby opening a window through which we can examine its effect on the regularity problem. 
To illustrate how this comes about, let us summarize the results which standard methods (H\"older and Sobolev inequalities) 
yield when attempting to estimate the rate of enstrophy production $\dot{D}_{1}$. The result in the unforced case is 
\bel{D1est1a}
\shalf\dot{D}_{1} \leq  \varpi_{0}\left(- D_{1}^{2}/4E + c\,D_{1}^{3}\right)\,,
\ee
where the dimensionless, bounded energy is $E = \nu^{-2}L^{-1}\I|\bu|^{2}dV$. This result has been known for a generation 
\cite{CF88,IDDS, DG95,FMRT01} and is derived for the reader in \S\ref{threereg}. As it stands, (\ref{D1est1a}) allows no control 
over $D_{1}$ beyond short times for arbitrarily large initial data or for long times from very small initial data. Moreover, dimensional 
scaling arguments suggest that no improvement on the $D_{1}^{3}$-term can be obtained when standard methods are used. However, 
\S\ref{threereg} shows that a re-working of this term by the insertion of the nonlinear depletion 
\bel{dep1b}
D_{m} \leq D_{1}^{A_{m,\lambda}}\,,\qquad\mbox{where}\qquad A_{m,\lambda} = \max_{t} A_{m}(t)
\ee
results in the $D_{1}^{3}$-term being replaced by one proportional to $D_{1}^{\xi_{m,\lambda}}$ (see (\ref{b9})) where\,:
\bel{ximdef}
\xi_{m,\lambda} = \frac{\chi_{m,\lambda}+2m-3}{2(m-1)}\,,\qquad\mbox{with}\qquad\chi_{m,\lambda} = A_{m,\lambda}(4m-3)\,.
\ee
The parameter $\lambda$, lying in the range $1\leq \lambda \leq 4$, appears through a scaling argument in \S\ref{amchoice} which 
suggests that $A_{m,\lambda}$ and $\chi_{m,\lambda}$ take the form
\bel{amlamdef}
A_{m,\lambda} = \frac{m\lambda + 1-\lambda}{4m-3}\quad\mbox{and}\quad\chi_{m,\lambda} = m\lambda + 1-\lambda\,.
\ee
Note that when $\lambda = 4$, then $A_{m,4} = 1$. The value of $\lambda$ chosen in the above range depends on the initial 
conditions of a given numerical simulation. Equations (\ref{ximdef}) and (\ref{amlamdef}) yield
\bel{ximdep}
\xi_{m,\lambda} = 1+\shalf\lambda
\ee
which is explicitly independent of $m$.  
To gain control over $D_{1}$, for long times and large initial data, it is thus necessary to restrict $\xi_{m,\lambda}$ to $\xi_{m,\lambda} < 2$ 
and $\lambda$ to the range\footnote{\sf The lower bound $\lambda \geq 1$ derives from the lower bound on $D_{m}$ in (\ref{dep1a}).} 
$1 \leq \lambda < 2$\,: see Fig. 1. It appears that the numerical data in \cite{DGGKPV13} can be fitted to (\ref{amlamdef}) with $\lambda$ 
sitting well within this range\,: $\lamin$ is chosen as the minimum value of $\lambda$ for any given numerical fit. In \S\ref{num} we suggest 
that the range $1.15 \leq \lamin \leq 1.5$ is appropriate for a range of initial conditions. 
%%%%%%%%%%%%%%%%%
\begin{figure}
\centering
\setlength{\unitlength}{5mm}
\begin{picture}(11,11)
\thicklines
\put(0,0){\vector(0,1){10}}
\put(0,0){\vector(1,0){15}}
%{\color{red}\qbezier(0,0)(4,2.5)(0,5)}
\thinlines
\put(0,10){\makebox(0,0)[b]{$D_{m}$}}
\thinlines
\put(15.7,-0.25){\makebox(0,0)[b]{$D_{1}$}}
\put(0,0){\line(1,1){10}}
\put(12,9.8){\makebox(0,0)[b]{\scriptsize $D_{m} = C_{m}D_{1}$}}
\qbezier(0,0)(6,6)(11,7)
\put(4,8){\makebox(0,0)[b]\textbf{\sf\scriptsize regime III}}
\put(8.3,7.8){\makebox(0,0)[b]\textbf{\sf\scriptsize regime II (weak)}}
\put(8.2,5.6){\makebox(0,0)[b]\textbf{\sf\scriptsize regime I (regular)}}
\put(6,2){\makebox(0,0)[b]\textbf{\sf\scriptsize not allowed}}
\put(12,6.9){\makebox(0,0)[b]{\scriptsize $\lambda  = 2$}}
\qbezier(0,0)(6,4)(11,4)
\put(12,3.8){\makebox(0,0)[b]{\scriptsize $\lambda  = 1$}}
\linethickness{.4mm}
{\qbezier[25](0,.2)(5,4.5)(9.7,5)}
{\qbezier[25](0,.2)(5,4.5)(11,4.7)}
\put(13.5,4.6){\makebox(0,0)[b]{\scriptsize $D_{m} = D_{1}^{A_{m,\lambda}}$}}
\end{picture}
\caption{\sf\scriptsize  A cartoon of the three regimes in the $D_{1}-D_{m}$ plane represented by the inequalities in 
(\ref{reg2}) at some value of $m > 1$, with $A_{m,\lambda}$ defined in (\ref{dep1b}). Two solid concave curves bound 
regime I\,: the lower curve derives from H\"older's inequality and corresponds to $\lambda=1$, whereas the upper curve 
is the upper limit of the regular regime (see \S\ref{1streg}) which corresponds to $\lambda=2$. The dotted curves 
approximately denote the region where the computations of \S\ref{num} lie for various values of $\lambda$ in the range 
$1.15 \leq \lambda \leq 1.5$.  The line $D_{m} = C_{m}D_{1}$ separates regimes II and III.}\label{phase}
\end{figure}
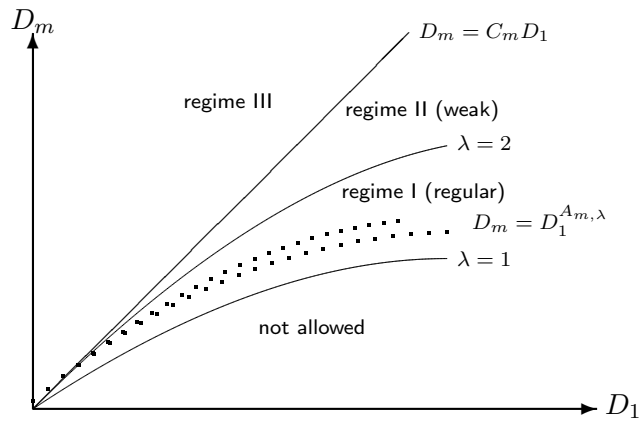
%%%%%%%%%%%
%\par\vspace{-2mm}%\noindent
\par\smallskip%noindent
While (\ref{dep1b}) is designated as regime I it is nevertheless theoretically possible that there exist other regimes beyond 
this (see Fig 1). The following regimes are defined and analyzed in \S\ref{1streg} and \S\ref{2ndreg}\,:
\beq{reg2}
D_{1}^{\alpha_{m}/2} &\leq& D_{m} \leq D_{1}^{A_{m,\lambda}}\,,\qquad\qquad~~\mbox{(regime~I)}\,;\nonumber\\
D_{1}^{A_{m,\lambda}}  &<& D_{m} \leq C_{m}D_{1}\,,~~\qquad\qquad\mbox{(regime~II)}\,;\\
C_{m}D_{1} &<& D_{m}\,,\qquad\qquad\qquad\qquad~~\mbox{(regime~III)}\,,\nonumber
\eeq
with regime I corresponding to the range $1\leq\lambda \leq 2$. 
The constant $C_{m}$ is determined in \S\ref{2ndreg}, where it is shown that regime II leads to no improvement in the $D_{1}^{3}$ 
estimate. Solutions are actually regular in regime III, but it is an open question whether this regime is physical. Fig. 1 is a cartoon of the 
regimes in (\ref{reg2}). 
\par\smallskip%\noindent
Remarkably, the two  respective values of the exponents $\xi_{m,\lambda} = 1+ \shalf\lambda$ and $\xi_{m,4}=3~(\lambda = 4)$ in 
regimes I and II, are close to those found in a paper by Lu and Doering \cite{LD08}, who used a numerical calculus of variations argument 
to find the value(s) of the exponent $\xi_{m,\lambda}$ when the rate of enstrophy production is maximized subject to the constraint 
$\mbox{div}\,\bu = 0$. They found that two branches existed, the lower being $D_{1}^{1.78}$ and the uppermost $D_{1}^{2.997}$. 
Later, Schumacher, Eckhardt and Doering \cite{SED10} suggested that 7/4 and 3 were the likely values of these two exponents\,; the 
exponent $\xi_{m,\lambda} = 7/4$ corresponds to $\lambda = 1.5$ which lies at the upper end of our observed range $1.15 \leq \lambda 
\leq 1.5$.
%%%%%%%%%%%%%%
\par\smallskip%\noindent
Boundedness from above of $D_{1}$ establishes existence and uniqueness and is the missing ingredient in the search for the existence of a 
global attractor $\mathcal{A}$  [10-13], albeit limited to regime I. In \S\ref{att} it is shown that estimates for the Lyapunov dimension of 
$\mathcal{A}$ are found to be (Proposition \ref{attdim})
\bel{introattdim}
d_{L}(\mathcal{A}) \leq c_{m}Re^{\frac{3(6-\lambda)}{5(2-\lambda)}}~~~\mbox{or}~~~
c_{m}Gr^{\frac{3(4-\lambda)}{5(2-\lambda)}}\,,
\ee
where $Re$ and $Gr$ are respectively the Reynolds and Grashof numbers defined in \S\ref{threereg}. \S\ref{spectra} shows that there 
is a corresponding energy spectrum in an inertial range for which $\mathcal{E}(k)\sim k^{-q_{m,\lambda}}$ where $q_{m,\lambda}= 
3-4/3\lambda$, with a cut-off at $L^{-1}Re^{3\lambda/4}$. The lower concave curve in Fig. 1 corresponds to $\lambda =1$ for which 
$q_{m,1} = 5/3$ with a cut-off at $L^{-1}Re^{3/4}$.  Regime I corresponds to $5/3 \leq q_{m,\lambda} < 7/3$.
\par\smallskip%\noindent
If these properties of regime I turn out to be typical of Navier-Stokes flows in periodic domains, then the existence and uniqueness results 
derived here are consistent with the observation that both numerical solutions [20-32] and experimental data \cite{Sreeni91,MS91}, while 
providing evidence of strong intermittency, have shown none of the violent super-exponential or singular growth observed in the 3D Euler 
equations \cite{HouLi06,RMK2013b}, nor have they shown any positive evidence of a lack of uniqueness. A related question is why a regime 
with such heavy depletion is favoured? Moreover, what vortical structures would correspond to it? Formally, using 
Sobolev and H\"older inequalities in $d$ dimensions ($1 \leq d \leq 3$) to estimate the vortex-stretching term, as in (\ref{D1est1a}), 
results in $\xi_{m,\lambda} = (6-d)/(4-d)$, which takes the expected value $\xi_{m,4}=3$ when $d=3$. The value of $d$ corresponding to 
$\xi_{m,\lambda} = 1+\shalf\lambda$ is $d_{\lambda} = 4(\lambda - 1)/\lambda$ which takes values from $d_{1.15}\approx 0.52$ to 
$d_{1.5} = 4/3$. This suggests that the dominant structures which give rise to the depletion observed in regime I could be the pasta-mix of 
tubes on which both vorticity and strain have long been numerically observed to accumulate \cite{Sreeni91,VM94} but also suggests that 
some vortical structures may lie closer to scattered points. 
%%%%%%%%%%%
\par\smallskip%\noindent
In contrast, \S\ref{2ndreg} shows that in regime II (labelled in Fig 1) these methods fail to find a proof of the existence of an attractor.  
Only Leray's weak solutions are known to exist and $q_{m,\lambda}$ lies at its outer limit with a value of $8/3$ for the sustenance of an 
energy cascade \cite{DG02,SF75}. In regime III, vorticity norms are under control, although it is possible that this regime represents an 
extreme state. While numerical evidence suggests that the Navier-Stokes equations operate in regime I only, it is still possible that solutions 
could jump between regimes, corresponding to some unusual initial conditions or higher Reynolds numbers. These possibilities are discussed 
in \S\ref{con}.

%%%%%%%%%%%%%%%%%%%%%
\section{\sf\large\textbf{Three Navier-Stokes regimes}}\label{threereg}

Consider the forced $3D$ Navier-Stokes equations on the periodic domain $[0,\,L]^{3}$\,: 
\bel{NSE1}
\partial_{t}\bu + \bu \cdot\nabla\bu = \nu\Delta\bu - \nabla p + \bdf(\bx)\,,
\ee
with $\mbox{div}\,\bu = 0$. The forcing function $\bdf(\bx)$ and its derivatives are considered 
to be $L^{2}$-bounded \cite{DF02}.  Estimates will be made in terms of the Grashof number $Gr$ and the Reynolds 
number whose definitions are \cite{DF02}
\beq{GRdef}
Gr &=& \frac{L^{3}f_{rms}}{\nu^2}\,,\qquad\qquad f_{rms}^2 = L^{-3}\|\bdf\|_{2}^{2}\,,\\
Re &=& \frac{LU_{0}}{\nu}\,,\qquad\qquad\quad U_{0}^{2} = L^{-3}\left<\|\bu\|_{2}^{2}\right>_{T}\,,
\eeq
and 
where the time average to time $T$ is given by 
\bel{timeav}
\left<g(\cdot)\right>_{T} = %\overline{\lim}_{g(0)}
\frac{1}{T}\int_{0}^{T}g(\tau)\,d\tau\,.
\ee
Doering and Foias \cite{DF02} have introduced a simplified form of forcing with the mild restriction that involves it 
peaking around a length scale $\ell$, which, for simplicity, is taken here to be the box length $L$. Then they have  
shown that Navier-Stokes solutions obey $Gr \leq c\,Re^{2}$ and that the global enstrophy satisfies
\bel{DF1}
\left<D_{1}\right>_{T} \leq GrRe + O\left(T^{-1}\right)\leq c\,Re^{3} + O\left(T^{-1}\right)\,.
\ee
In fact, all the $\left<D_{m}\right>_{T}$ for $1\leq m\leq \infty$ are bounded \cite{JDGCMS11}.

%%%%%%%%%%%%%%
\subsection{\sf\textbf{A summary of numerical work}}\label{num}

The results from several numerical experiments, some of which were reported in \cite{DGGKPV13}), are summarized in Figs. 2, 3 
and 4 which show plots of
\bel{amdef}
A_{m}(t) = \frac{\ln D_{m}(t)}{\ln D_{1}(t)}
\ee
versus time $t$, with the exception of Fig. 3c, in which the horizontal axis is $Re_{\lambda}$\,: this the conventional notation for 
the Taylor micro-scale Reynolds number so the subscript $\lambda$ should not be confused with the parameter $\lambda$ in (\ref{dep1b}). 
The $D_{m}$ are replaced by the time averages $\left<D_{m}\right>_{T}$ and the $A_{m}$ by $\overline{A}_{m}$\,:
%%%%%%%%%%
\ben\itemsep -1mm

\item Figs. 2a-d come from a pseudo-spectral $512^{3}$ simulation of the forced Navier-Stokes equations on a $(2\pi)^{3}$ 
domain with random initial conditions\,: in all cases $\max_{t}A_{m} \leq 0.46$, with $m=2$ at the upper limit, but with values 
dropping close to about $0.37$ as $m\to9$. Fig. 2a is the result of Kolmogorov forcing $f(x,y,z) = f_{0} \sin(k_{1}x)$ with $f_{0} 
= 0.005$ and $k_{1}=1$, which keeps the Grashof number $Gr$ constant ($Gr = 8.8\times 10^{7}$). Figs. 2b-d are the result of 
white-noise forcing restricted to those modes for which $|\bk| = 1$, i.e.,
\bel{wn1} %  ($Gr_{1} = 124025\,;~Gr_{2} = 248050\,;~Gr_{3} = 1240251$) 
f(x,y,z) = f_{0}(t)\cos \{k_{1} x + k_{2}y + k_{3}z\}\,.
\ee
The amplitude $f_0 (t)$ is a zero-mean ($Gr = L^{3} f_{A}\nu^{-2}$), Gaussian white noise with variance 
$\left< f_{0}(t)\,f_{0}(t')\right> = f_{A}\,\delta (t - t')$.  The values of $Re_{\lambda}$ for the simulations shown in Figs. 
2b-d are $Re_{\lambda_{1,2,3}}= 97, 117$ and $192$ respectively. 

\item Fig. 3a is a decaying simulation of fully developed Navier-Stokes turbulence performed by Kerr \cite{RMK2012a,DGGKPV13} 
who used an anisotropic $1024\times2048\times 512$ mesh in a $2\pi (2\times8\times1)$ domain, with symmetries applied to the 
$y$ and $z$ directions. As summarized in \cite{DGGKPV13,RMK2012a}, the simulation has long anti-parallel vortices as initial 
conditions from which develop three sets of reconnections at $t=16,\,96$ and $256$.  The figure is a plot of $A_{m}$ for $m=2$ 
descending to $m=9$ where $\max_{t}A_{m}$ takes its maximum at $m=2$ (0.46), and decreases to about $3/8$ as $m\to9$. 

\item Fig. 3b shows a plot from a decaying version of the simulation in Figs. 2a-d. $\max_{t}A_{m} \leq 0.43$, but decreases close to 
$0.37$ as $m\to9$. 

\item Fig. 3c derives from a DNS data-base using a massively parallel pseudo-spectral code run on $10^{5}$ processors, which 
includes simulations with resolutions up to $4096^{3}$ and Taylor-Reynolds number up to $Re_{\lambda} \sim 1000$ 
\cite{DYS2008,DY2010,YDS2012}. In order to maintain a stationary state, turbulence is forced numerically at the large scales. 
Results are shown using the stochastic forcing of Eswaran \& Pope \cite{EP88} (denoted as EP), as well as a deterministic 
scheme described in \cite{DY2010} (denoted as FEK). The figure shows the $m=2$ case descending to $m=6$\,: open and 
closed symbols in the figure correspond to EP and FEK forcing, respectively. These schemes are summarized in more detail in 
\cite{DGGKPV13}.  Here $\overline{A}_{m}$ is defined by $\overline{A}_{m} = \ln \left<D_{m}\right>_{T}/\ln \left<D_{1}
\right>_{T}$ while the horizontal axis denotes values of $Re_{\lambda}$ which goes up to $10^{3}$, while $\max\overline{A}_{m} 
\leq 0.42$.
%%%%

\item The simulations above have been performed in the range $2 \leq m \geq 9$. In Fig. 4 we give one example of a simulation in 
the range $1 \leq m \leq 2$. Three values of $\left(m,\,A_{m,\lambda}\right)$ are given in table 2. There it can be seen that the 
range of $\lambda$ is $1.19 \leq \lambda \leq 1.5$. 

\een
%%%%%%%% Figs 2a, b, c, d %%%%%%%%%
\begin{figure}[ht]\label{fig2}
\centering
\includegraphics[scale=.30]{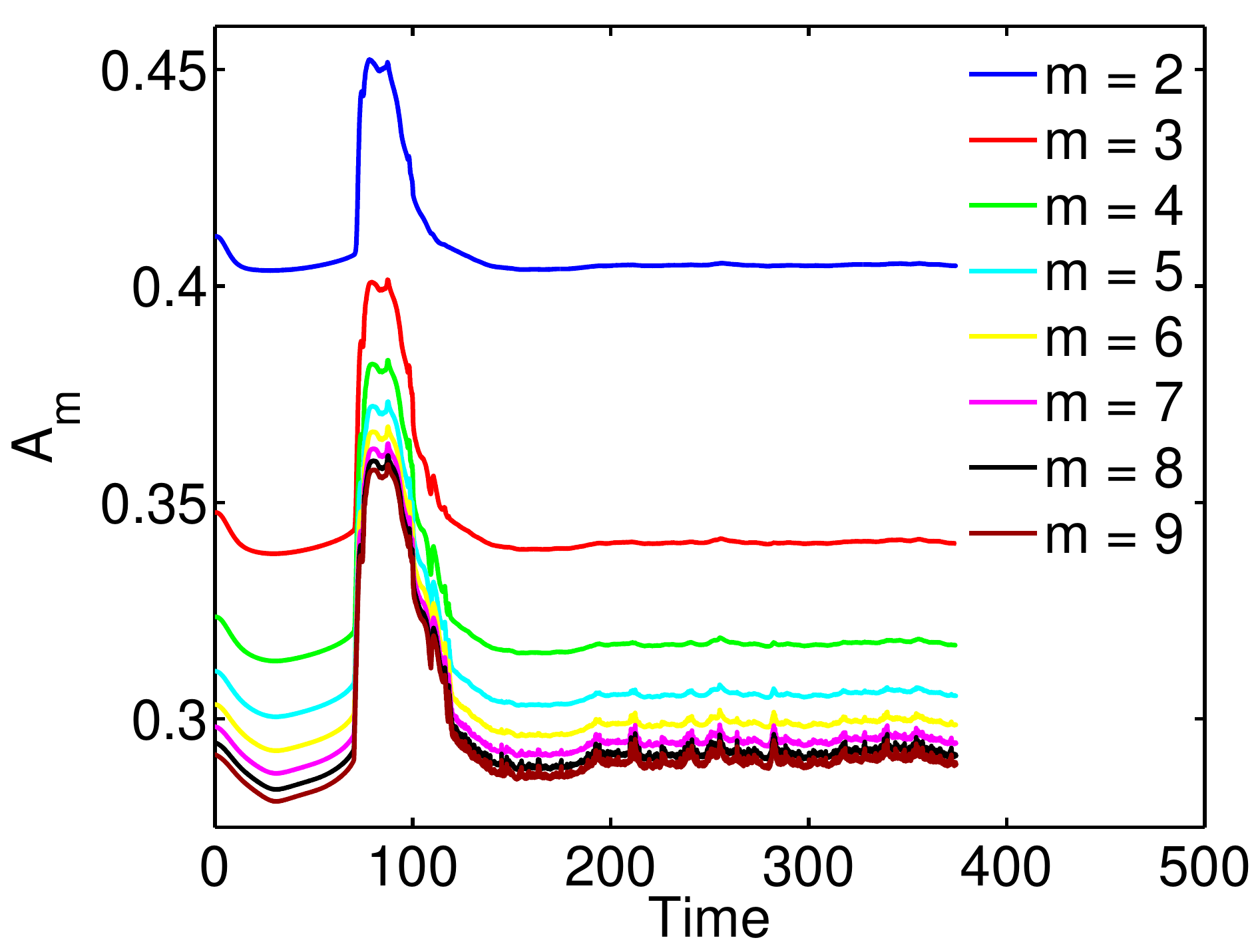}
\includegraphics[scale=.30]{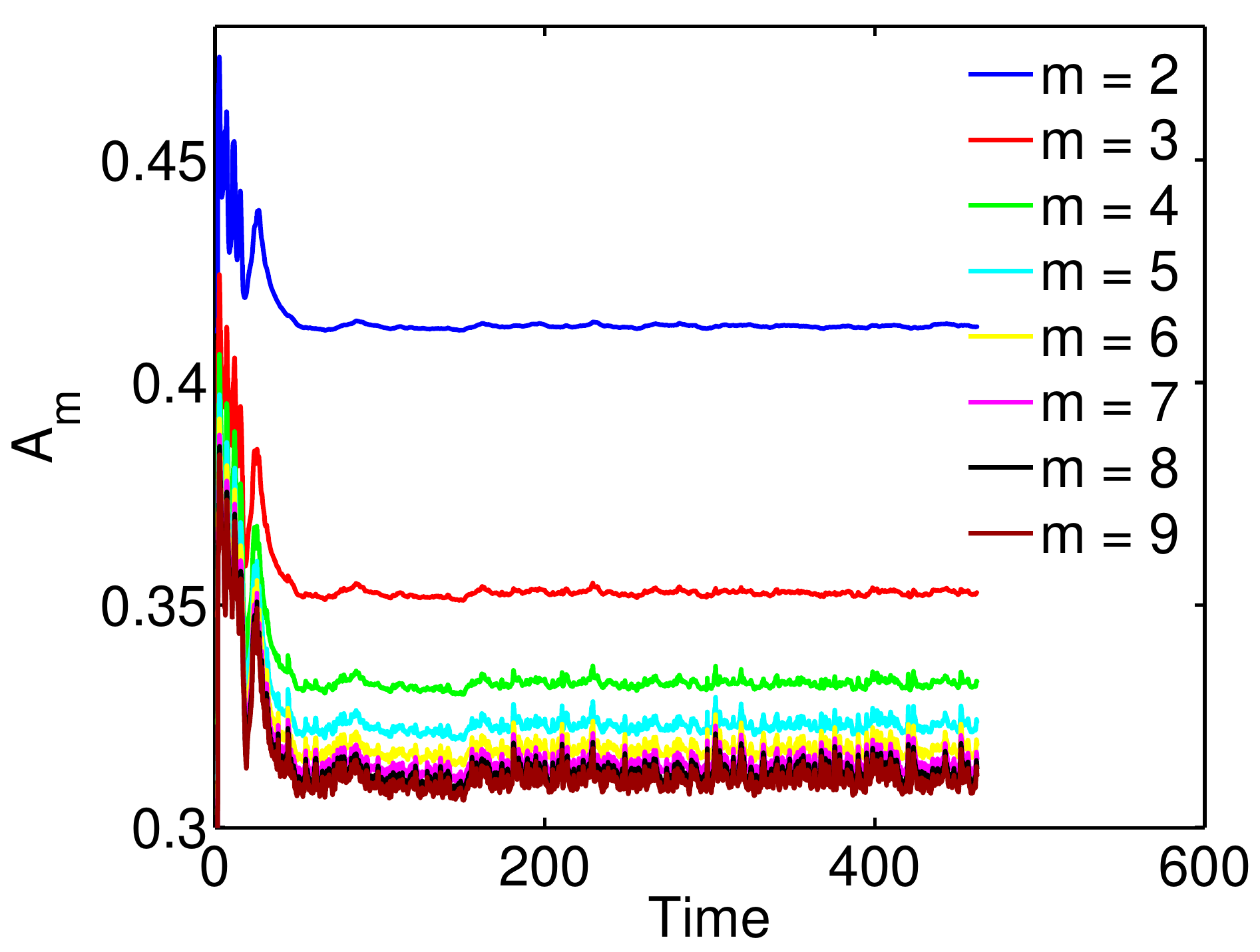}\qquad\\
\includegraphics[scale=.30]{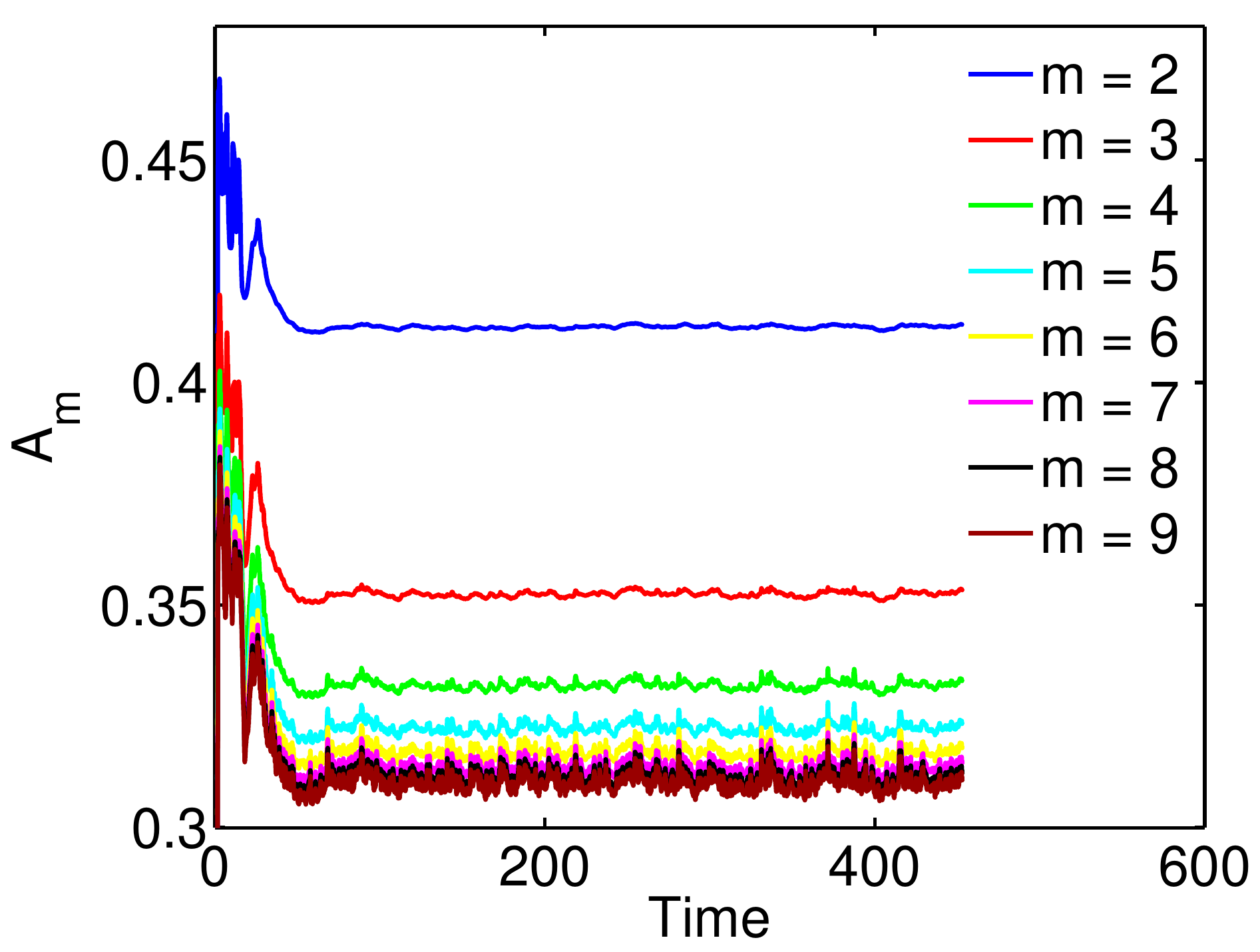}
\includegraphics[scale=.30]{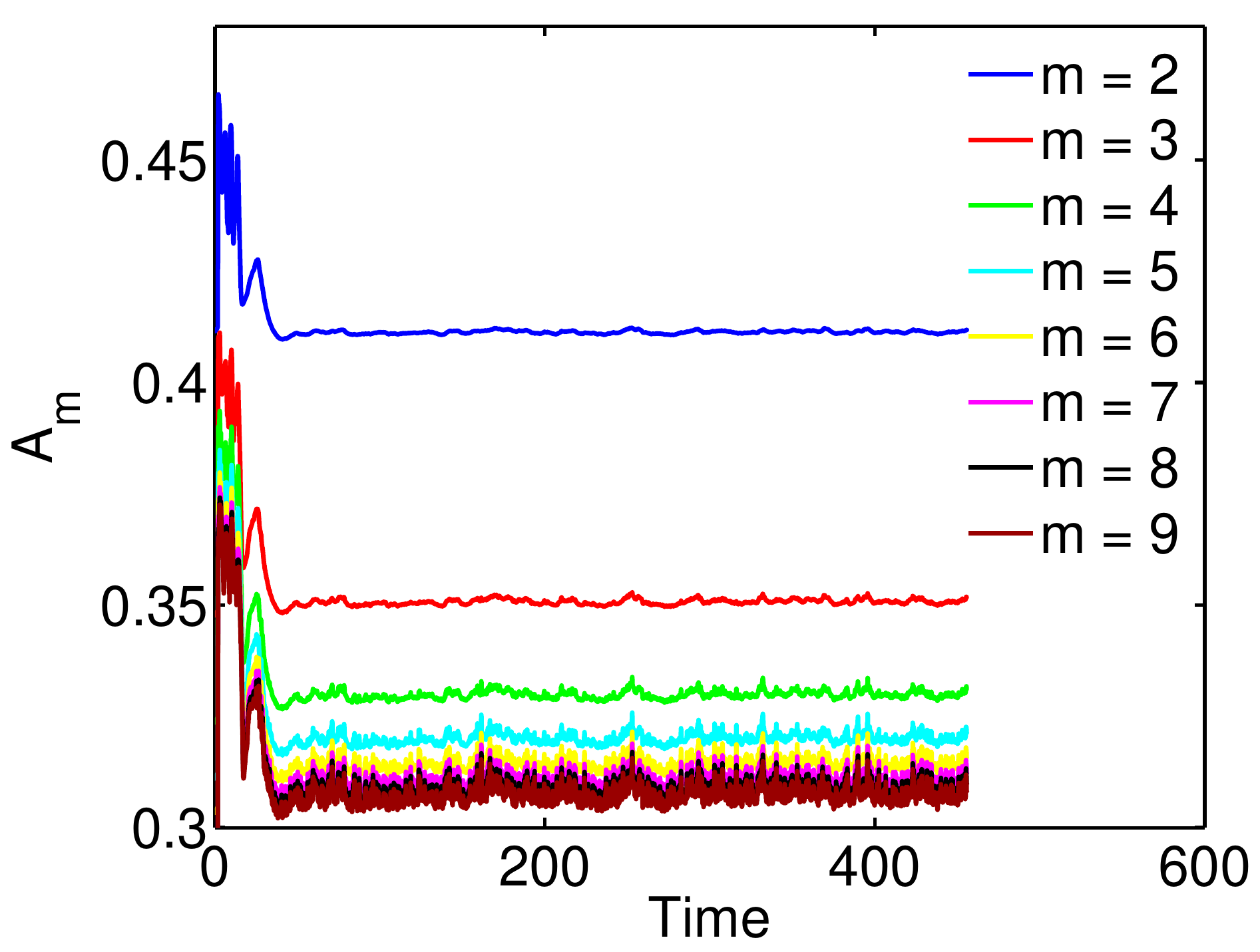}
\caption{\sf\scriptsize Plots of $A_{m}$ versus time for four forced simulations\,: in all of these $\max_{t} A_{m} < 0.5$. Figs 2a-d
are the result of pseudo-spectral $512^{3}$ simulations on a cubical $[0,\,2\pi]^{3}$-domain with random initial conditions. 
Fig 2a is the result of Kolmogorov forcing with $|\bdf_{0}|=0.005$ and $k_{1}=1$, while figs 2b-d are plots for three different 
values of $Re_{\lambda}$, namely $Re_{\lambda_{1,2,3}} = 97, 117$ and $192$ respectively, under the influence of white-noise 
forcing\,: see the text for a more detailed explanation.}
\end{figure} % $Gr_{1} = 124025\,,~Gr_{2} = 248050$, and $Gr_{3} = 1240251$)
%%%%%%%%%%%%%%%% end Figs 2a,b,c,d %%%%%%%%%%%
%%%
%%%%%%%%%%%%%%%% Figs 3a,b,c %%%%%%%%%%%%%
\begin{figure}[ht]\label{fig3}
\centering
\includegraphics[scale=.25]{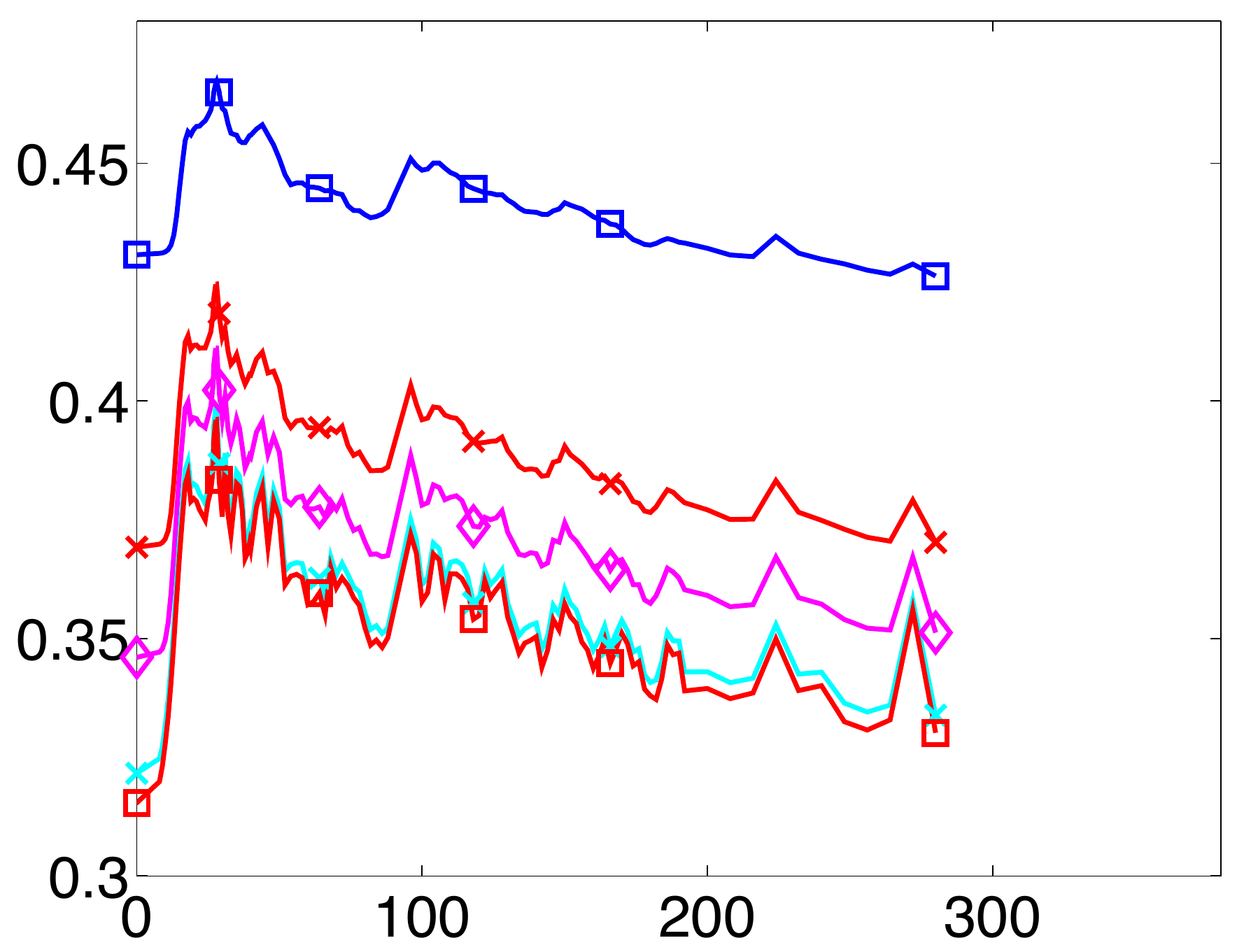}
\includegraphics[scale=.24]{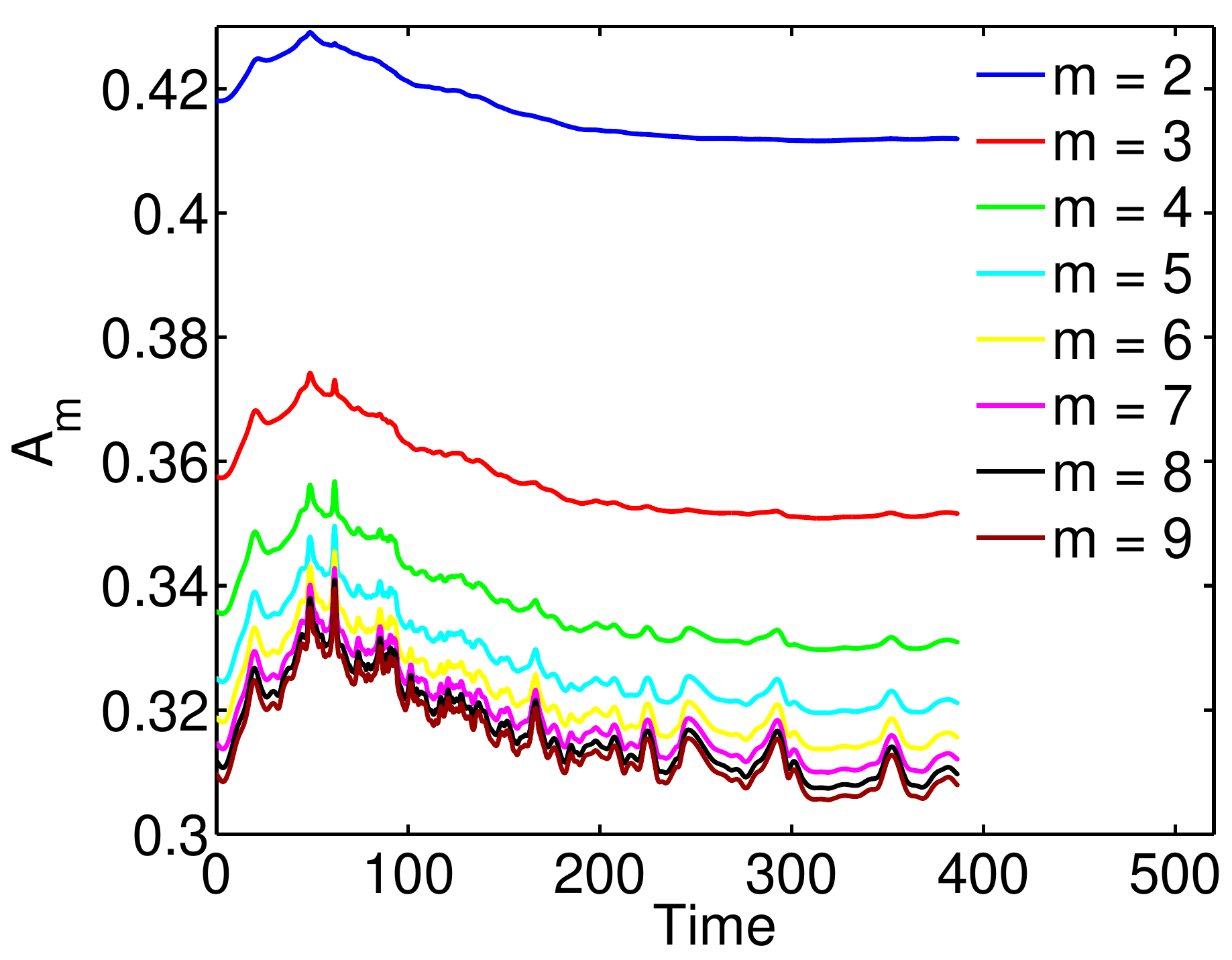}\quad
\includegraphics[scale=.25]{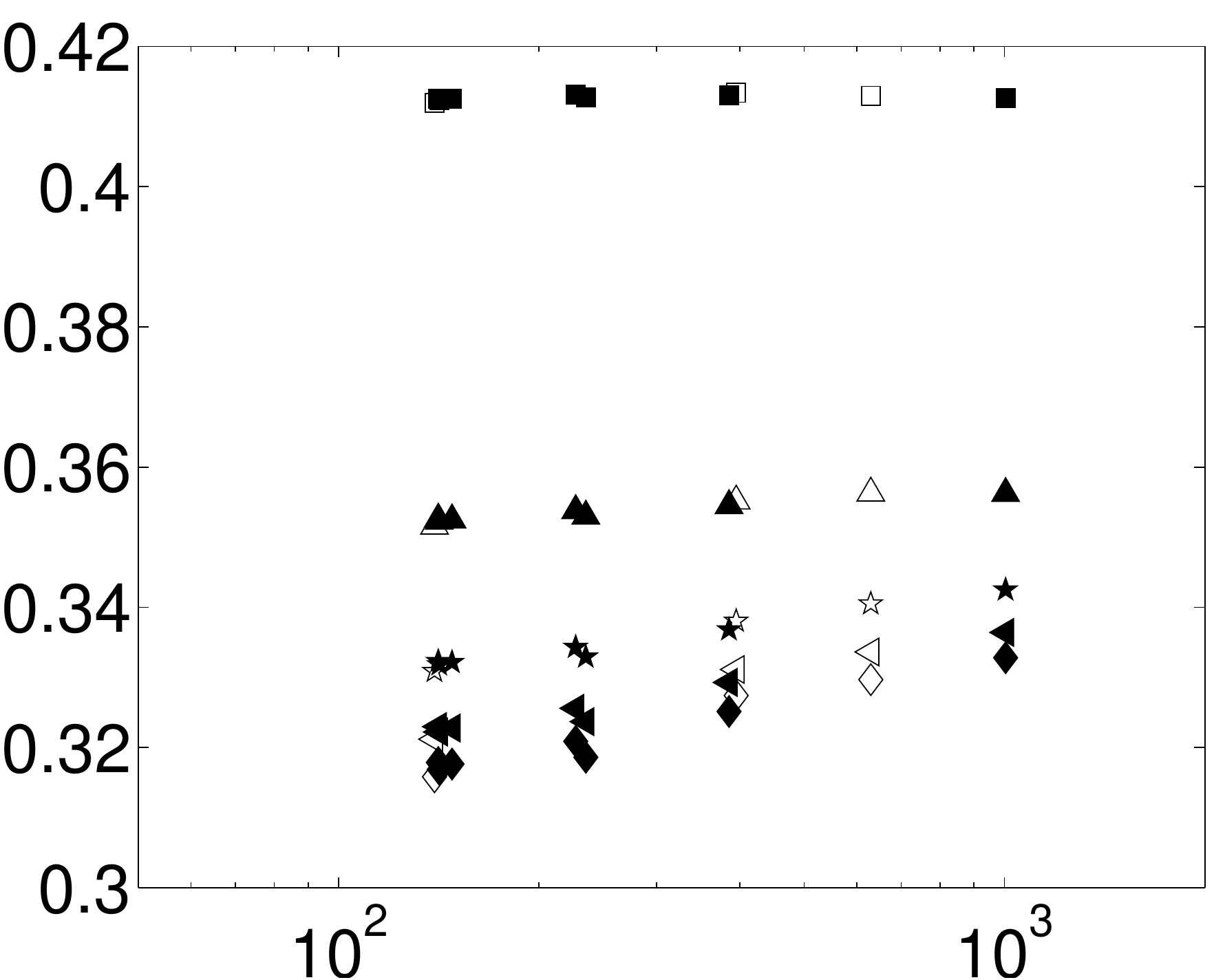}
\caption{\sf\scriptsize Plots 3a,\,b are of $A_{m}$ defined in (\ref{amdef}) versus time for two decaying simulations for which 
$\max_{t}A_{m} < 0.5$. 
Fig. 3a is a $1024\times2048\times512$ pseudo-spectral simulation on a long $4\pi\times16\pi\times2\pi$ domain with anti-parallel initial 
conditions \cite{RMK2012a,DGGKPV13}. Fig. 3b is a decaying version of the $512^{3}$ simulation as in Figs. 2a-d. Fig. 3c is a plot of 
$\overline{A}_{m}$ defined by $\overline{A}_{m} = \ln \left<D_{m}\right>_{T}/\ln \left<D_{1}\right>_{T}$ arising from the TAMU 
database with the horizontal axis denoting values of $Re_{\lambda}$. Open and closed symbols denote results from two types of 
forcing (EP and FEK) for statistically steady flows\,: see \cite{DGGKPV13}.}
\end{figure}
\par\noindent
%%%%%%%%%%%%%%%% end Figs 3a,b,c %%%%%%%%%%%
%\par\vspace{-5mm}
%%%%%%%%%%%%%% %%% Fig 4 %%%%%%%%%%%%%
\begin{figure}[ht]\label{fig4}
\centering
\includegraphics[scale=.33]{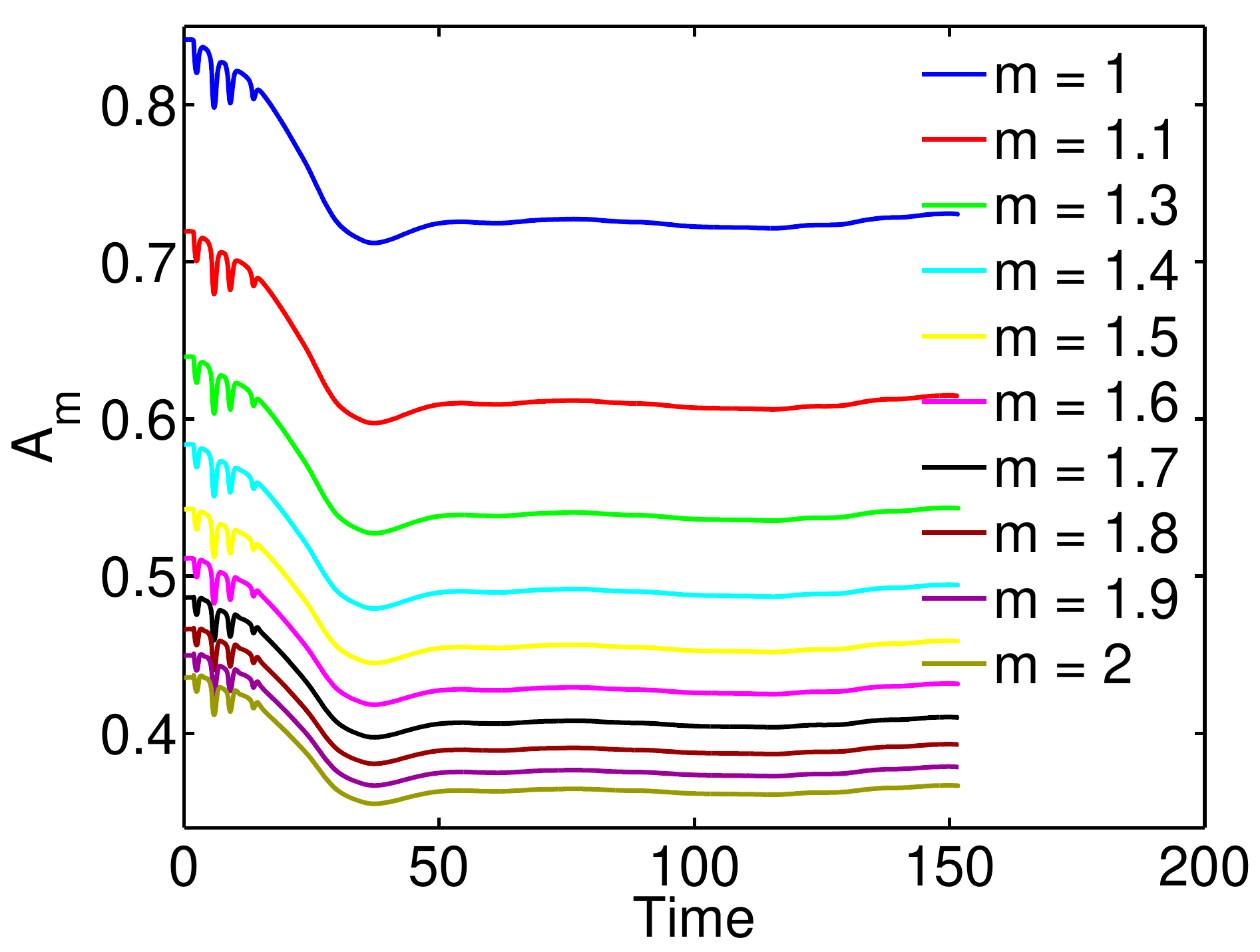}
\hspace{5mm}
\includegraphics[scale=.33]{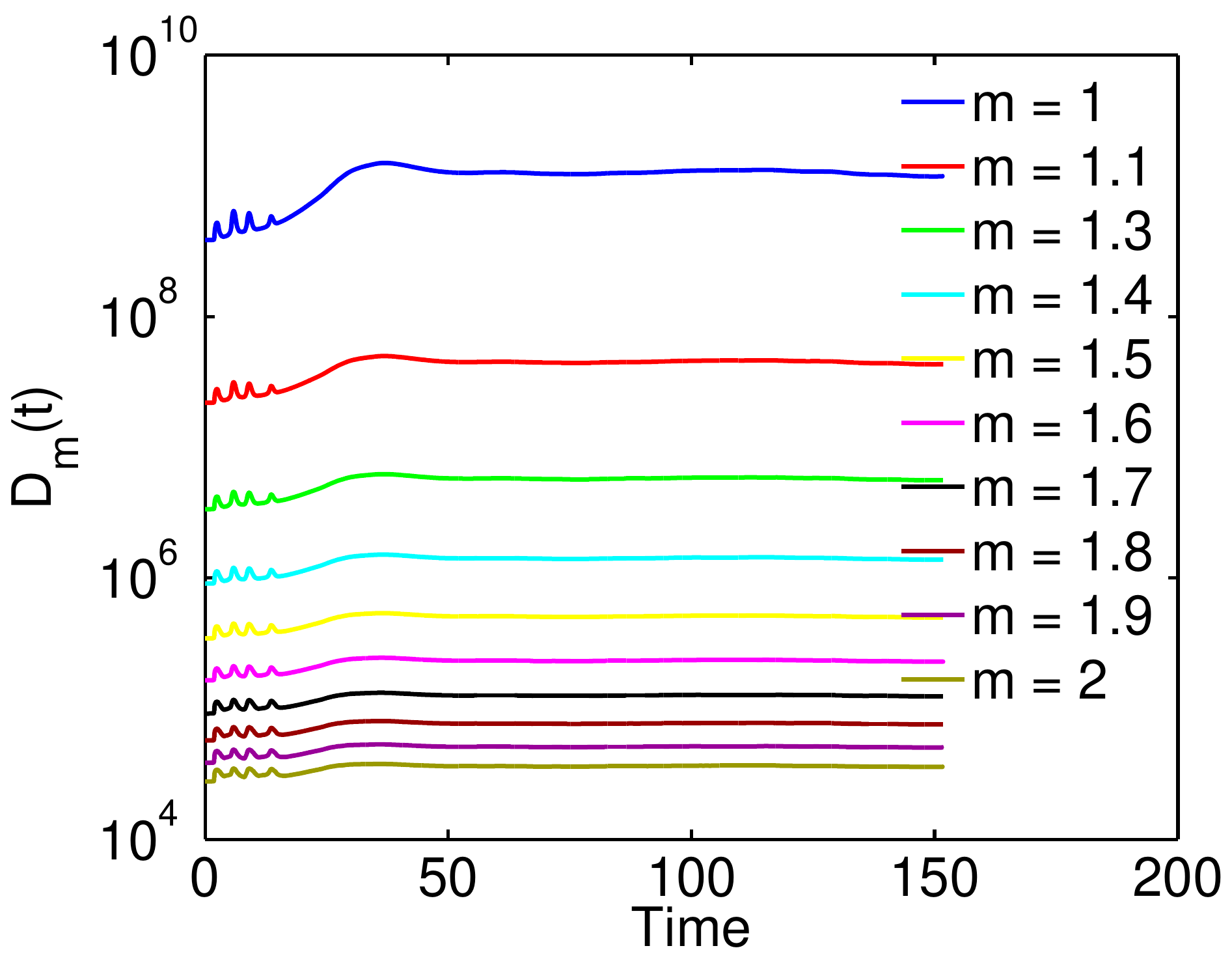}
%%%%%%%%%%%%
\caption{\sf\scriptsize Example of a simulation with white-noise forcing, as in Fig. 2d ($Re_{\lambda}=192$) performed 
in the range $1 \leq m \leq 2$. In the right-hand figure $D_{1}$ is included whereas it is not in Fig. 2d.}
\end{figure}
%%%%%%%%%%% end Fig 4 %%%%%%%%%%%

%%%%%%%%
\subsection{\sf\textbf{How to choose $\max_{t} A_{m}$}}\label{amchoice}

%%%%%%%%%%%%%%
\begin{table}[h]\label{amtab}
\begin{center}
\begin{tabular}{||c||c|c|c|c|c|c|c|c|c||}
\hline
$\lambda$ & 1.1 & 1.15  & 1.2  & 1.25 & 1.3 & 1.35 & 1.4 & 1.45 & 1.5\\\hline
$A_{2,\lambda}$    & 0.42 & 0.43 & 0.44 & 0.45& 0.46 & 0.47 & 0.48 & 0.49 & 0.5\\\hline
$A_{6,\lambda}$    & 0.31 & 0.32 & 0.33 & 0.35& 0.36 & 0.36 & 0.38 & 0.39 & 0.40\\\hline
$A_{9,\lambda}$    & 0.30 & 0.31 & 0.32 & 0.33& 0.35 & 0.36 & 0.37 & 0.38 & 0.39\\\hline
\end{tabular}
\caption{\sf\scriptsize Table of values of $\lambda$ and $A_{2,\lambda},~A_{6,\lambda}$ and $A_{9,\lambda}$ 
corresponding to Figs. 2 and 3.}
\end{center}
\end{table}
%\par\vspace{-5mm}\noindent
The numerical experiments reported above show that $A_{m}$ has values lying in a wide range. Is there a way of choosing $\max_{t}A_{m}$ 
as a function of $m$ in a simple manner consistent with the results of these simulations? The following is a consistency argument based on 
the inequalities the $D_{m}$ must obey. 
\par\smallskip%\noindent
Firstly it is easy to prove that 
\bel{ap1}
\Omega_{m}^{m(p+q)} \leq \Omega_{m-p}^{q(m-p)}\Omega_{m+q}^{p(m+q)}\,.
\ee
Let $p=m-1$ and $q=1$ to give $\Omega_{m}^{m^2} \leq \Omega_{1}\Omega_{m+1}^{m^{2}-1}$. In terms of $D_{m}$, this 
translates to 
\bel{ap3}
D_{m}^{m^2} \leq D_{1}^{\alpha_{m}/2}D_{m+1}^{\alpha_{m}(m^{2}-1)/\alpha_{m+1}}\,.
\ee
Suppressing the $\lambda$-label on $A_{m+1}$ in the depletion $D_{m+1} \leq D_{1}^{A_{m+1}}$, we obtain
\bel{ap4}
D_{m}\leq D_{1}^{\frac{1+A_{m+1}(m-1)(4m+1)}{m(4m-3)}}\,.
\ee
For the exponent on the right-hand side of (\ref{ap3}) to be consistent with $D_{m} \leq D_{1}^{A_{m}}$, we require 
\bel{ap5}
\frac{1+A_{m+1}(m-1)(4m+1)}{m(4m-3)} = A_{m}\,.
\ee
By using the definition $\chi_{m} = A_{m}(4m-3)$, this reduces to
\bel{ap6}
1+ (m-1)\chi_{m+1} = m\chi_{m}\,,
\ee
which is solved to give\,:
\begin{proposition}\label{deltaprop}
The solution of (\ref{ap6}) is given by
\bel{ap7}
\chi_{m,\lambda} = m\lambda  + 1-\lambda\,,\qquad\mbox{or}\qquad 
A_{m,\lambda} = \frac{m\lambda + 1-\lambda}{4m-3}\,,
\ee
where the constant $\lambda$ lies in the range $1 \leq \lambda \leq 4$.
\end{proposition}
%%%%%%%%%%%%%%%%%%%%%%%%%%%%%
The fit of (\ref{ap7}) to the figures in \S\ref{num} is not perfect in the sense that numerical trajectories do not 
follow \textit{exactly} the concave curves of Fig. 1, so the appropriate value of $\lambda$ from an initial condition 
needs to be estimated.  To achieve this, we label as $\lambda_{m}$ those values computed from $\max_{t}A_{m}$ 
in a given figure. These increase slightly with $m$\,; for example, in Fig 3a, $\lambda_{2} =1.4$ at $m=2$, whereas 
$\lambda_{9}=1.45$ at $m=9$ (see table 1). $\lamin$, defined in (\ref{lamindef}), is taken as the minimum 
of a set of values of $\lambda$ computed over a range of $m$ from a given initial condition. This can then be used in 
the estimates for the attractor dimension or energy spectrum in the following sections.
%%%%%%%%%%%%%%%%%%%
\bel{lamindef}
\lamin = \min \left\{\lambda_{m}\,: 2 \leq m \leq N\right\}\,.
\ee
%\par\vspace{-5mm}\noindent
\begin{table}[h]\label{mlesstab}
\begin{center}
\begin{tabular}{||c||c|c|c||}
\hline
$m$ & 1.1 & 1.5  & 1.9 \\\hline
$A_{m,\lambda}$ & 0.85 & 0.55 & 0.45\\\hline
$\lambda$ & 1.5 & 1.3 & 1.19\\\hline
\end{tabular}
\caption{\sf\scriptsize Table of values of $\lambda$ corresponding to $A_{1.1,\lambda},~A_{1.5,\lambda}$ and 
$A_{1.9,\lambda}$ corresponding to Fig. 4.}
\end{center}
\end{table}

%%%%%%%%%%%%%%%%%%%%%%%%
\subsection{\sf\textbf{A division into three regimes}}

The scaling \textit{ansatz} for $A_{m,\lambda}$ in (\ref{ap7}) derived in Proposition 1 suggests that the $D_{1}-D_{m}$ plane 
can be divided into different regimes for the range\footnote{\sf The two regimes I and II could be merged by using the range $1\leq 
\lambda\leq 4$ but this leaves a gap between $D_{1}$ and $C_{m}D_{1}$ which causes technical difficulties.} $1\leq\lambda\leq 2$\,:
\par\vspace{2mm}\noindent
\textbf{Regime I\,:} $D_{1}^{\frac{m}{4m-3}} \leq D_{m} \leq D_{1}^{A_{m,\lambda}}$\,: the lower bound is $\Omega_{1} 
\leq \Omega_{m}$ expressed in the $D_{m}$-notation.
\par\vspace{2mm}\noindent
\textbf{Regime II\,:} $D_{1}^{A_{m,\lambda}} < D_{m} \leq  C_{m} D_{1}$, where\footnote{\sf The constants $c_{i,m}$ in $\varpi_{1,m} 
= \varpi_{0}c_{1,m}^{-1}$ and $\varpi_{2,m} = \varpi_{0}c_{2,m}$ have the following properties \cite{JDGPRS10,JDGJMP12}\,:  
$c_{1,m}$ is a Sobolev constant multiplied by $m^{2}/(m-1)$ whereas $c_{2,m}$ derives from the constant in $\|\nabla\bu\|_{p} 
\leq c_{p}\|\bom\|_{p}$ for $1 < p < \infty$.} $C_{m}^{\eta_{m} } = \varpi_{2,m}/\varpi_{1,m} = c_{1,m}c_{2,m}$, and 
$\eta_{m} = \frac{2m}{3(m-1)}$.
\par\vspace{2mm}\noindent
\textbf{Regime III\,:} $ C_{m} D_{1} < D_{m}$.
\par\vspace{2mm}\noindent
Regime III appears in the following way. Using the standard contradiction method\footnote{\sf This method assumes the existence of a maximal 
interval of existence and uniqueness on an interval $[0,\,T^{*})$, which means that $D_{1}$ must be infinite at $T^*$\,: then, in any subsequent 
calculation, one considers the behaviour of $D_{1}$ as $t\to T^*$. If this limit is finite then a contradiction has occurred, thus invalidating the original 
assumption of a \textit{maximal} interval. This cannot be zero so it must be infinite. The value of the method is that it allows the differentiation 
of the $D_{m}$ on $[0,\,T^{*})$.}, 
for $1 < m < \infty$ ($\varpi_{i,m}$ are constants), the $D_{m}$ obey the differential inequality \cite{JDGJMP12,JDGiutam13} 
\beq{s3}
\dot{D}_{m} &\leq& D_{m}^{3}\left\{-\varpi_{1,m}\left(\frac{D_{m+1}}{D_{m}}\right)^{\rho_{m}}  
+ \varpi_{2,m}\right\} + \varpi_{3,m}Gr D_{m}^{\frac{\alpha_{m}-1}{\alpha_{m}}}\,,\\
\rho_{m} &=& \twothirds m(4m+1)\label{rhomdef}\,.
\eeq 
Moreover, for $m > 1$ it is easily proved that 
\bel{s4}
\frac{D_{m}}{D_{1}}\leq \left(\frac{D_{m+1}}{D_{m}}\right)^{(m-1)(4m+1)}\,,
\ee
which changes (\ref{s3}) and (\ref{rhomdef}) into
\beq{s5}
\dot{D}_{m} &\leq& D_{m}^{3}\left\{-\varpi_{1,m}\left(\frac{D_{m}}{D_{1}}\right)^{\eta_{m}}  + 
\varpi_{2,m}\right\} + \varpi_{3,m}Gr D_{m}^{\frac{\alpha_{m}-1}{\alpha_{m}}}\,,\\
\eta_{m} &=& \frac{2m}{3(m-1)}\label{delmdef}\,.
\eeq
Clearly, in regime III the combination of terms within the braces is negative and can be neglected. In this regime 
the dissipation is sufficiently strong to control solutions rather than depletion reducing the nonlinearity. In the 
unforced case, the $D_{m}$ always decay\,; at most, they grow only algebraically in time in the forced case (see 
\S\ref{2ndreg}). Moreover, $\Omega_{1} \leq \Omega_{m}$ universally implies that
\bel{A1}
\left(\frac{D_{m}}{D_{1}}\right)^{\eta_{m}}D_{m}^{2} \geq 1\,,
\ee
which means that $D_{m} \geq 1$ in regime I, while in regime II there is a lower bound $D_{m}^{2} \geq 
C_{m}^{-\eta_{m}}$.

%%%%%%%%%%%%%%%%%%%%%%%%%%%%%%%%%%%
\section{\sf\large\textbf{Regime I}}\label{1streg}

%%%%%%%%%%%%%%%%%%%%%%%%%%%%
\subsection{\sf\large\textbf{Depletion resulting in an absorbing ball for $D_{1}$}}\label{ball}

It has long been understood that the $H_{1}$-norm of the velocity field ($D_{1}$ in the notation of this paper) controls 
all the regularity 
properties of $3D$ Navier-Stokes equations \cite{CF88,IDDS,DG95,FMRT01}. It is also the essential missing ingredient in 
the search for the proof of the existence of a $3D$ Navier-Stokes global attractor. What is required is an ``absorbing ball'' for 
this norm, which consists of a ball of finite radius into which all solutions are drawn for large times. In what follows, estimates 
are made for the forced case in terms of the Grashof number $Gr$ or Reynolds number $Re$. In the unforced case the conclusions 
regarding the finiteness of $D_{1}$ still stand except that the radius of the ball decays and the attractor is just the origin.
%Without the existence of this ball 
%it cannot be proved that the $L^2$-ball for the velocity field $\bu$ is compact, which is a necessary condition for the 
%existence of a global attractor \cite{CF88,IDDS,DG95,FMRT01}.
\par\smallskip%\noindent
In this context it is difficult to handle a wide variety of forcing functions analytically. For simplicity we shall remain with the 
properties of the forcing as in Doering and Foias \cite{DF02} who took forcing at a single scale $\ell$, taken here to be the 
box-scale $L$, to make estimates in terms of the Grashof number $Gr$ or Reynolds number $Re$ defined in (\ref{GRdef}). 
\par\smallskip%\noindent
The first task is to illustrate why the standard estimate for $D_{1}$ produces an apparently unsurmountable problem. Note 
that from the definition of the $D_{m}$ in (\ref{s1}) $D_{1} = L\nu^{-2}\|\bom\|_{2}^{2}$
%\qquad\qquad D_{2} = \left(\nu^{-1}L^{5/4} \|\bom\|_{4}\right)^{4/5}\,.
so, using the standard contradiction method (see footnote 10), a formal differential inequality for $D_1$ is
\bel{b2}
\shalf \dot{D}_{1} \leq L\nu^{-2}\left\{-\nu \I |\nabla\bom|^{2}\,dV + \I |\nabla\bu||\bom|^{2}\,dV + 
L^{-1}\left(\I|\bom|^{2}\,dV\right)^{1/2}\|\bdf\|_{2}\right\}\,.
\ee 
Dealing with the negative term first, an integration by parts gives
\bel{b3}
\I |\bom|^{2}\,dV \leq \left(\I|\nabla\bom|^{2}dV\right)^{1/2}\left(\I|\bu|^{2}dV\right)^{1/2}\,,
\ee
where the dimensionless energy $E$ is defined as
\bel{b4}
E = \nu^{-2}L^{-1}\I|\bu|^{2}\,dV\,.
\ee
which is always bounded such that [10-13] %\cite{CF88,IDDS,DG95,FMRT01}
\bel{b7}
\overline{\lim}_{t\to\infty}E \leq c\,Gr^{2}\,.
\ee
Then the nonlinear term in (\ref{b2}) can be estimated in two ways\,: 
\ben\itemsep -1mm
\item By using a Sobolev inequality in the standard way [10-13]\,; %\cite{CF88,IDDS,DG95,FMRT01}\,; 

\item By invoking the nonlinear depletion of regime I.  
\een
(1) The standard method simply involves a Schwarz inequality to estimate the nonlinear term as
\bel{b5a}
\I |\nabla\bu||\bom|^{2}\,dV \leq \|\bom\|_{2}\|\bom\|_{4}^{2}\,.
\ee
After the application of the Sobolev inequality $\|\bom\|_{4} \leq c\,\|\nabla\bom\|_{2}^{3/4}\|\bom\|_{2}^{1/4}$, 
this becomes
\beq{b5b}
\I |\nabla\bu||\bom|^{2}\,dV &\leq& c\,\|\nabla\bom\|_{2}^{3/2}\|\bom\|_{2}^{3/2}\nonumber\\
&\leq&
\frac{3\nu}{4}\|\nabla\bom\|_{2}^{2} + \frac{c}{4\nu^{3}}\|\bom\|_{2}^{6}\,.
\eeq
(\ref{b2}) then becomes
\bel{b6}
\shalf \dot{D}_{1} \leq \varpi_{0}\left(-\frac{1}{4}\frac{D_{1}^{2}}{E} + c\,D_{1}^{3} + Gr D_{1}^{1/2}\right)\,.
\ee
Clearly the cubic nonlinearity is too strong for the quadratic negative term\,: all we can deduce is that $D_{1}$ is bounded from 
above only for short times or for small initial data. The difficulty caused by this term has been known for many decades\,: see 
\cite{CF88,IDDS,DG95,FMRT01} and also Lu and Doering \cite{LD08}. 
\par\medskip\noindent
(2) Now we turn to using the nonlinear depletion of regime I. How might the insertion of $D_{m} \leq D_{1}^{A_{m,\lambda}}$ mollify 
the cubic exponent in (\ref{b6})? We return to (\ref{b2}) and estimate the nonlinear term as
\beq{b8a}
\I |\nabla\bu||\bom|^{2}dV &=&\I |\bom|^{\frac{2m-3}{m-1}}|\bom|^{\frac{1}{m-1}}|\nabla\bu|dV\nonumber\\
&\leq & \left(\I|\bom|^{2}dV\right)^{\frac{2m-3}{2(m-1)}}
\left(\I|\bom|^{2m}dV\right)^{\frac{1}{2m(m-1)}}\left(\I|\nabla\bu|^{2m}dV\right)^{\frac{1}{2m}}\nonumber\\
&\leq & c_{m}\left(\I|\bom|^{2}dV\right)^{\frac{2m-3}{2(m-1)}}\left(\I|\bom|^{2m}dV\right)^{\frac{1}{2(m-1)}}\nonumber\\
&=& c_{m}L^{3}\varpi_{0}^{3} D_{1}^{\frac{2m-3}{2m-2}}D_{m}^{\frac{4m-3}{2m-2}}\,,\qquad\qquad 1 < m < \infty\,.
\eeq
based on $\|\nabla\bu\|_{p} \leq c_{p} \|\bom\|_{p}$, for $1 < p < \infty$. Inserting the depletion $D_{m} \leq D_{1}^{A_{m,\lambda}}$,
\bel{b8b}
L\nu^{-2}\I |\nabla\bu||\bom|^{2}\,dV \leq c_{m}\varpi_{0}D_{1}^{\xi_{m,\lambda}}\,,
\ee
where $\xi_{m,\lambda}$ is defined as in (\ref{ximdef}) but repeated here
\bel{ximdefA}
\xi_{m,\lambda} = \frac{\chi_{m,\lambda}+2m - 3}{2(m-1)}\,,\qquad\qquad \chi_{m,\lambda} = 
A_{m,\lambda}(4m-3) = m\lambda +1 -\lambda\,.
\ee
Thus we have
\bel{ximdeldef}
\xi_{m,\lambda} = 1+ \shalf \lambda\,,
\ee
\textbf{which is explicitly $m$-independent.} Thus the equivalent of (\ref{b6}) is
\bel{b9}\boxed{
\shalf \dot{D}_{1} \leq \varpi_{0}\left(-\frac{D_{1}^{2}}{E} + c_{m} D_{1}^{1+ \shalf\lambda} + Gr D_{1}^{1/2}\right)\,.}
\ee
Given that $E$ is bounded above, $D_{1}$ is always under control provided $\lambda$ is restricted to the range $1 \leq \lambda< 2$. 
This is expressed in the following\,:
%%%%%%%%%%%%%%
\begin{proposition}\label{prop2}
If the solution always remains in regime I ($1 \leq \lambda < 2$), there exists an absorbing ball for $D_{1}$ of radius
\bel{D1ball}
\overline{\lim}_{t\to\infty} D_{1}\leq c_{m}Gr^{\frac{4}{2 - \lambda}} + O\left(Gr^{4/3}\right)\,.
\ee
\end{proposition}
\textbf{Remark 1\,:} The range of control over $D_{1}$ in $1 \leq \lambda < 2$ can be extended to $\lambda = 2$ as (\ref{b9}) 
shows that there is an exponentially growing bound on $D_{1}$ at this value.
\par\smallskip\noindent
\textbf{Remark 2\,:} Note that the values of $\lambda = \lamin$ corresponding to the numerical experiments in \S\ref{threereg} 
lie well within the range ($1 \leq \lambda < 2$) of validity, as illustrated by Fig. 1.
\par\smallskip\noindent
\textbf{Remark 3\,:} From (\ref{D1ball}) and the standard properties of the Navier-Stokes equations \cite{CF88,IDDS,DG95,FMRT01}, 
we conclude that a global attractor $\mathcal{A}$ exists in this regime, which is a compact $L^{2}$-bounded ball for the 
velocity field $\bu$. $c_{m}$ is a generic constant dependent only on $m$.

%%%%%%%%%%%%%%%%%%%%%%%%%%%%%%%%%%%%

\subsection{\sf\large\textbf{An estimate for the attractor dimension}}\label{att}

It is now possible to estimate the Lyapunov dimension of the global attractor $\mathcal{A}$, which has been 
shown to exist as a result of Proposition \ref{prop2}, subject to the depletion in regime I. A connection 
between the system dynamics and the 
attractor dimension is provided by the notion of the Lyapunov exponents through the Kaplan-Yorke formula.  
For ODEs the Lyapunov exponents control the exponential growth or contraction of volume elements in phase 
space\,: the Kaplan-Yorke formula expresses the balance between volume growth and contraction realized 
on the attractor.  It has been rigorously applied to global attractors in PDEs by Constantin and Foias 
\cite{CF85,CF88}\,: see also \cite{IDDS,DG95,GT97}.  The formula is the following\,: for Lyapunov exponents 
labelled in descending order and designated by $\mu_{n}$, the Lyapunov dimension $d_{L}$ is defined in 
terms of these by 
\bel{dldef}
d_{L} = N -1 + \frac{\mu_{1}+\ldots + \mu_{N-1}}{-\mu_{N}},
\end{equation}            
where the number $N$ of $\mu_{n}$ is chosen to satisfy
\bel{KY2}
\sum_{n=1}^{N-1}\mu_{n}\geq 0\hspace{1cm}\mbox{but}\hspace{1cm}\sum_{n=1}^{N}\mu_{n} < 0\,.
\ee
Note that according to the definition of $N$, the ratio of exponents in (\ref{dldef}) satisfies
\bel{KY1}
0 \leq \frac{\mu_{1}+\ldots + \mu_{N-1}}{-\mu_{N}} < 1\,,
\ee          
so the formula generally yields a non-integer dimension such that 
\bel{KY3}
N - 1 \leq d_{L} < N\,. 
\ee
The value of $N$ that turns the sign of the sum of the Lyapunov exponents, as in (\ref{KY2}), is that value 
of $N$ that bounds above $d_{L}$ and hence the Hausdorff and fractal dimensions $d_{H}$ and $d_{F}$. 
For a discussion of generalized dimensions see the paper by Hentschel and Procaccia \cite{HP83}. To use the 
method for PDEs as developed in \cite{CF85,CF88} the phase space is replaced by $\bu\in L^{2}\cap\mbox{div}
\,\bu=0$, which is infinite dimensional. The solution $\bu(t)$ forms an orbit in this space, with different sets of 
initial conditions $\bu(0)+\delta\bu_{i}(0)$, which evolve into $\bu(t)+\delta\bu_{i}(t)$ for $i =1,\ldots,N$.  
The linearized form of the Navier-Stokes equations in terms of $\delta\bu$ of $\bu$ is
\bel{NS2}
\partial_{t}(\delta\bu) + \bu\cdot\nabla\delta\bu + \delta\bu\cdot\nabla\bu  = \nu \Delta\delta\bu -\nabla\delta p\,,
\ee
which can also be written in the form 
\bel{trace1}
\partial_{t}(\delta\bu) = \mathcal{M}\delta\bu\,.
\ee
If they are chosen to be linearly independent, initially these $\delta\bu_{i}$ form an $N$-volume or parallelpiped of volume
\bel{vol1}
V_{N}(t) = \left|\delta\bu_{1}\wedge\delta\bu_{2}\ldots\wedge\delta\bu_{N}\right|\,.
\ee
It is now necessary to find the time evolution of $V_{N}$.  This is given by 
\bel{vol2}
\dot{V}_{N} = V_{N}Tr \left[\mathbf{P}_{N}\mathcal{M}\mathbf{P}_{N}\right]\,,
\ee
which is easily solved to give  
\bel{trace2} 
V_{N}(t) = V_{N}(0) \exp \int_{0}^{t}Tr \left[\mathbf{P}_{N}\mathcal{M}\mathbf{P}_{N}\right](\tau )\,d\tau\,. 
\ee
$\mathbf{P}_{N}(t)$ is an $L^{2}$-orthogonal projection, using the orthonormal set of functions $\{\bphi_{i}\}$, onto the 
finite dimensional subspace $\mathbf{P}_{N}L^{2}$, which spans the set of vectors $\delta\bu_{i}$ for $i=1,\,...,\,N$. 
%\bel{span}\mbox{span}\left\{\delta_{m}\bu_{1},\delta_{m}\bu_{2},\ldots,\delta_{m}\bu_{N}\right\}.\ee
In terms of the time average $\left<\cdot\right>_{t}$ up to time $t$, the sum of the first $N$ global Lyapunov
exponents is taken to be \cite{CF85,CF88}
\bel{global1}
\sum_{n=1}^{N}\mu_{n} = \left< Tr \left[\mathbf{P}_{N}\mathcal{M}\mathbf{P}_{N}\right]\right>_{t} \,.
\ee
As in (\ref{KY2}), we want to find the value of $N$ that turns the sign of 
$\left<Tr\left[\mathbf{P}_{N}\mathcal{M}\mathbf{P}_{N}\right]\right>_{t}$
and for which volume elements contract to zero. This value of $N$ bounds above 
$d_{L}$ as in (\ref{dldef}). To estimate this we write
\bel{tr1}
Tr \left[\mathbf{P}_{N}{\cal M}\mathbf{P}_{N}\right]  = \sum_{n=1}^{N}
\I\bphi_{n}\cdot\left\{\nu\Delta\bphi_{n} 
- \bu\cdot\nabla\bphi_{n} - \bphi_{n}\cdot\nabla\bu 
- \nabla\tilde{p}\left(\bphi_{n}\right)\right\}\,dV.
\ee 
Since $\mbox{div}\,\delta_{m}\mbox{\boldmath$u$}_{n} = 0$ for all $n$, then 
$\mbox{div}\,\mbox{\boldmath$\phi$}_{n}=0$ also and so the pressure term
integrates away, as does the second term
\bel{tr2}
Tr \left[\mathbf{P}_{N}{\cal M}\mathbf{P}_{N}\right] 
\leq - \nu \sum_{n=1}^{N} \I |\nabla\mbox{\boldmath$\phi$}_{n}|^{2}\,dV
+ \sum_{n=1}^{N} \I |\nabla\bu|\,|\mbox{\boldmath$\phi$}_{n}|^{2}\,dV .
\ee
Because the $\mbox{\boldmath$\phi$}_{n}$ are orthonormal they obey the relations
\bel{tr3}
\sum_{n=1}^{N} \I|\mbox{\boldmath$\phi$}_{n}|^{2}\,dV = N,
\hspace{1cm}\mbox{and}\hspace{1cm}
Tr \left[\mathbf{P}_{N}(-\Delta)\mathbf{P}_{N}\right] 
= \sum_{n=1}^{N} \I|\nabla\mbox{\boldmath$\phi$}_{n}|^{2}\,dV.
\ee
% Notice also that since the $\delta_{m}\mbox{\boldmath$u$}_{i}$ solve the linear parabolic PDE (\ref{trace1}), 
% then from the smoothing effect, this means that $\delta_{m}\mbox{\boldmath$u$}_{i}\in H^{2}(\Omega )$.  
%The $\mbox{\boldmath$\phi$}_{n}$ have some orthogonality properties of which we can take advantage 
%to estimate the relative magnitudes of the negative and positive terms on the left hand side. 
In $3D$ the $\mbox{\boldmath$\phi$}_{n}$ satisfy the Lieb-Thirring inequalities \cite{CF85,CF88,FMRT01,IDDS} 
for orthonormal functions
\bel{LT1}
\I\left(\sum_{n=1}^{N}
|\mbox{\boldmath$\phi$}_{n}|^{2}\right)^{5/3}\!\!dV
\leq c\,\sum_{n=1}^{N} \I |\nabla\mbox{\boldmath$\phi$}_{n}|^{2}\,dV,
\ee
where $c$ is independent of $N$.  Moreover, it is known that the first $N$ eigenvalues of the 
Stokes operator in three-dimensions satisfy
\bel{lower1}
Tr\left[\mathbf{P}_{N}(-\Delta )\mathbf{P}_{N}\right] \geq c\,N^{5/3}L^{-2}\,.
\ee
To exploit the Lieb-Thirring inequality (\ref{LT1}) to estimate the last term in (\ref{tr2}) we write it as 
\bel{nltA}
\sum_{n=1}^{N} \I |\nabla \bu|\,|\mbox{\boldmath$\phi$}_{n}|^{2}\,dV
\leq \left[\I|\nabla \bu |^{5/2}\,dV\right]^{2/5}\,\left[\I
\left(\sum_{n=1}^{N}|\mbox{\boldmath$\phi$}_{n}|^{2}\right)^{5/3}\!dV\right]^{3/5}.
\ee
Hence, using (\ref{LT1}) and time averaging $\left<\cdot\right>_{t}$, we find
\begin{eqnarray}\label{nlt3}
\left<\sum_{n=1}^{N} \I|\nabla \bu |\,|\mbox{\boldmath$\phi$}_{n}|^{2}\,dV\right>_{t}
& \leq & c\,\left<\left(Tr \left[\mathbf{P}_{N}(-\Delta)\mathbf{P}_{N}
\right]\right)^{3/5}\left(\I|\nabla\bu|^{5/2}\,dV
\right)^{2/5}\right>_{t}\nonumber\\
& \leq & \frac{3\nu}{5}\left<Tr \left[\mathbf{P}_{N}(-\Delta)\mathbf{P}_{N} \right]\right>_{t}
+ \frac{2c}{5\nu^{3/2}}\left<\int_{\Omega}|\nabla\bu|^{5/2}\,dV\right>_{t}
\end{eqnarray}
and so (\ref{tr2}) can be written as 
\bel{tr4}
\left<Tr \left[\mathbf{P}_{N}{\cal M}\mathbf{P}_{N}\right]\right>_{t}
\leq - \frac{2}{5}\nu\left<Tr \left[\mathbf{P}_{N}(-\Delta)\mathbf{P}_{N}\right]\right>_{t} + \frac{2}{5}
c\,\nu^{-3/2}\left<\I |\nabla \bu|^{5/2}\,dV\right>_{t}\,.
\ee
%%%%%%%%%%%
% \bel{ad1}Tr \left[\mathbf{P}_{N}{\cal M}\mathbf{P}_{N}\right]  \leq - c_{1}\nu N^{5/3} + 
% L^{2}\nu^{-3/2}\left<\|\nabla\bu\|_{5/2}^{5/2}\right> \ee
%%%%%%%%%%%
To estimate the nonlinear term we use H\"older's inequality to obtain ($m > 1$)
\beq{ad2a}
\I\left|\nabla\bu\right|^{5/2}\,dV &\leq& c\, \I |\bom|^{5/2}\,dV\non\\
&\leq& c\,\left(\I|\bom|^{2}\,dV\right)^{\frac{4m-5}{4(m-1)}}\left(\I|\bom|^{2m}\,dV\right)^{\frac{1}{4(m-1)}}\nonumber\\
&\leq& c\, \varpi_{0}^{5/2}L^{3}D_{1}^{\frac{4m-5}{4(m-1)}}D_{m}^{\frac{4m-3}{4(m-1)}}\,.
\eeq 
%Thus, for $m=2$ \bel{ad2b}
%\I\left|\nabla\bu\right|^{5/2}\,dV \leq c\,\varpi_{0}^{5/2}L^{3}D_{1}^{3/4}D_{2}^{5/4}\,.\ee
Therefore, using this and (\ref{lower1}), we find
\bel{ad3a}
\varpi_{0}^{-1}\left<Tr \left[\mathbf{P}_{N}{\cal M}\mathbf{P}_{N}\right]\right>_{t}  \leq 
- c_{1}N^{5/3} + c_{2}\left<D_{1}^{\frac{4m-5}{4(m-1)}}D_{m}^{\frac{4m-3}{4(m-1)}}\right>_{t}\,.
\ee
It is at this point where the depletion of nonlinearity $D_{m} \leq D_{1}^{A_{m,\lambda}}$ is used, thereby giving
%\bel{ad3b}\left<D_{1}^{3/4}D_{2}^{5/4}\right>  \leq c\,\left<D_{1}^{11/8}\right> \leq c\,%\left<D_{1}\right>\left(\overline{\lim}_{t\to\infty}D_{1}\right)^{3/8}\ee
\bel{b15}
\left<D_{1}^{\frac{4m-5}{4(m-1)}}D_{m}^{\frac{4m-3}{4(m-1)}}\right>_{t} \leq \left<D_{1}\right>_{t} 
\left(\overline{\lim}_{t\to\infty}D_{1}\right)^{\frac{\chi_{m,\lambda}-1}{4(m-1)}}\,,
\ee
where $\chi_{m,\lambda} = A_{m,\lambda}(4m-3)$ as defined in (\ref{ximdef}). 
%%%%%%%%%%%
%%%%%%%
Proposition \ref{prop2} and the estimate $\left<D_{1}\right>_{t} \leq c\, GrRe$ from \cite{DF02} then allow us to write
\bel{ad4a}
\left<D_{1}\right>_{t} \left(\overline{\lim}_{t\to\infty}D_{1}\right)^{\frac{\chi_{m,\lambda}-1}{4(m-1)}}
\leq c\,(Gr Re) Gr^{\frac{\chi_{m,\lambda}-1}{2m-1 - \chi_{m,\lambda}}} 
\leq c\, Re^{\frac{6m - 5 - \chi_{m,\lambda}}{2m-1-\chi_{m,\lambda}}}\,,
\ee
and so (\ref{ad3a}) can be written as 
\bel{ad4b}
\left<Tr \left[\mathbf{P}_{N}{\cal M}\mathbf{P}_{N}\right] \right>_{t} \leq  \varpi_{0}\left(- c_{1}N^{5/3} + 
c_{2}Re^{\frac{6m - 5 - \chi_{m,\lambda}}{2m-1-\chi_{m,\lambda}}}\right)\,.
\ee
To find an estimate solely in terms of $Gr$ the $(Gr Re)$-term of (\ref{ad4a}) is replaced by $Gr^{2}$. Choosing 
$\chi_{m,\lambda}$ as in (\ref{ap7}), we have proved\,:
%%%%%%%%%%%%%%%%%%%
\begin{proposition}\label{attdim}
If the solution always remains in regime I the Lyapunov dimension of the global attractor $\mathcal{A}$ is estimated as
\bel{ad5a}
d_{L}(\mathcal{A}) \leq c_{1,m}Re^{\frac{3}{5}\left(\frac{6 - \lambda}{2-\lambda}\right)}\,,
\ee
or, alternatively, as 
\bel{ad5b}
d_{L}(\mathcal{A}) \leq c_{2,m}Gr^{\frac{3}{5}\left(\frac{4-\lambda}{2-\lambda}\right)}\,.
\ee
\end{proposition}

%%%%%%%%%%%%%%%%%%%%%%%%%%

\section{\sf\large\textbf{Regimes II and III}}\label{2ndreg}

In \S\ref{1streg}, regime I has been defined to lie in the region $D_{m} \leq D_{1}^{A_{m,\lambda}}$  for $1 \leq \lambda < 2$, with 
regime II defined as the region where this inequality has been reversed up to $C_{m}D_{1}$. One could fuse regimes I and II together 
by taking $\lambda$ in the wider range $1 \leq \lambda \leq 4$ but we have no control over $D_{1}$ for $2 < \lambda \leq 4$. In this 
section we choose to remain with the definition of regime II as in (\ref{reg2}).
\par\smallskip%\noindent
To test whether there is any depletion in regime II let us repeat inequality (\ref{b8a}) for the nonlinear term and use $D_{m} 
\leq C_{m} D_{1}$
\beq{sec1}
\I |\nabla\bu||\bom|^{2}\,dV &\leq& c\,L^{3}\varpi_{0}^{3} D_{1}^{\frac{2m-3}{2m-2}}D_{m}^{\frac{4m-3}{2m-2}}\non\\
 &\leq& c\,L^{3}\varpi_{0}^{3}C_{m}^{\frac{4m-3}{2m-2}} D_{1}^{3}\,.
\eeq
Thus $\xi_{m,4}=3$ and there is no depletion of nonlinearity in the upper bound $D_{m} \leq C_{m} D_{1}$. Moreover, when the scaling 
argument in \S\ref{amchoice} is repeated, this too shows no depletion.  To test whether the dissipation term in (\ref{s5}) is changed by 
the use of the lower bound $D_{1}^{A_{m,\lambda}} < D_{m}$ we consider first (\ref{A1}) 
\bel{sec3a}
D_{m}^{2}\left(\frac{D_{m}}{D_{1}}\right)^{\eta_{m}} > D_{1}^{A_{m,\lambda}(2+\eta_{m}) - \eta_{m}} \equiv 
D_{1}^{\Delta_{m,\lambda}}\,,
\ee
which improves the lower bound of unity in (\ref{A1}) and thereby increases the dissipation.  In fact 
\bel{Deltalam}
\Delta_{m,\lambda} = \frac{2}{3}(\lambda - 1)\,.
\ee
Let us assume that initial data is placed in regime II at a time $t_{0}$\,: then dividing (\ref{s5}) by 
$D_{m}^{3}$ we find 
\bel{sec4}
\shalf \frac{d~}{dt}D_{m}^{-2} \geq D_{m}^{-2}\left\{\varpi_{1,m}D_{1}^{\Delta_{m,\lambda}}\right\} - \varpi_{2,m}\,,
\ee
where, for convenience, we have taken the unforced case \cite{JDGJMP12}. An integration over $[t_{0},\,t]$ gives 
\bel{sec5}
e^{-(t-t_{0})g(t)}D_{m}^{-2}(t) \geq D_{m}^{-2}(t_{0}) -
2\varpi_{2,m}\int_{t_{0}}^{t}e^{-(\tau-t_{0})g(\tau)}\,d\tau
\ee
where
\bel{gdef1}
g(t) = \frac{ 2 \varpi_{1,m}}{t-t_{0}}\int_{t_{0}}^{t}D_{1}^{\Delta_{m,\lambda}}\,d\tau\,.
\ee
The main question here is whether there exists a sufficiently large \textit{lower} bound on the time average $g(t)$ to prove that the right 
hand side of (\ref{sec5}) never develops a zero for some wide range of initial data? The problem is that the size of $\int_{t_{0}}^{t}
D_{1}^{\Delta_{m,\lambda}}d\tau$ over \textit{very short intervals} $[t_{0},\,t]$ is indeterminate.  This lower bound would have to be 
large enough \textit{on arbitrarily small intervals} for the negative integral of the exponential in (\ref{sec5}) to be always smaller than 
$D_{m}^{-2}(t_{0})$ to prevent a zero forming on the right-hand side. 
\par\smallskip%\noindent
Finally, regime III is easily dealt with because the condition $C_{m} D_{1} < D_{m}$ allows us to drop two of the three terms in 
the (\ref{s5}) leaving us with $\dot{D}_{m} \leq 0$ in the unforced case, thus implying decay from initial data. In the forced case 
$\dot{D}_{m} \leq \varpi_{3,m}Gr D_{m}^{1-1/\alpha_{m}}$ and so it follows that any $D_{m}$ that satisfies this is bounded for 
all time as in 
\bel{sec7}
D_{m} \leq \left[D_{m}^{\alpha_{m}^{-1}}(t_{0}) + \alpha_{m}^{-1}\varpi_{3,m}Gr\left(t - t_{0}\right)\right]^{\alpha_{m}}
\ee

%%%%%%%%%%%%%%%
\section{\sf\large\textbf{Energy spectra and typical length scales in regimes I \& II}}\label{spectra}

Some ideas are explained in this section on how information might be extracted from the analysis on the properties of an 
energy spectrum $\mathcal{E}(k)$ corresponding to regimes I and II.  Doering and Gibbon \cite{DG02} have shown how 
to associate bounds of time averages with the moments of this spectrum by following some ideas in \cite{Frisch95,SF75}. 
It is these arguments we shall summarize first. 
\par\smallskip%\noindent
In the standard manner, we define 
\bel{Hndef}
H_{n}(t) = \I |\nabla^{n}\bu|^{2}\,dV\qquad\mbox{with}\qquad H_{0} = \I |\bu|^{2}\,dV\,,
\ee
where the label $n$ refers to derivatives. Then it was shown in \cite{DG02} that to take proper account of the forcing 
these require an additive adjustment such that
\bel{Fndef}
F_{n} = H_{n} + \tau^{2}\|\nabla^{n}\bdf\|_{2}^{2}\,,
\ee
where $\tau^{-1} \sim \varpi_{0}Gr^{\shalf+ \varepsilon}$ for any $\varepsilon > 0$. This formalism now allows us to 
define the set of `wave-numbers' $\kappa_{n,0}$ and $\kappa_{n,1}$ such that 
\bel{kappadef}
\kappa_{n,0}^{2n} = F_{n}/F_{0}\,,\qquad\qquad \kappa_{n,1}^{2(n-1)} = F_{n}/F_{1}\,.
\ee
Using the fact that 
\bel{H1}
\shalf\dot{H}_{1} \leq -\nu H_{2}+ \I |\nabla\bu||\bom|^{2}dV + \mbox{forcing}\,,
\ee
which is just another way of expressing (\ref{b2}), we can re-visit the inequality in (\ref{b8a}) to estimate the integral in (\ref{H1})
with the application of the depletion of regime I
\bel{H2a}
\I |\nabla\bu||\bom|^{2}dV \leq \varpi_{0}\left(L^{3}\varpi_{0}^{2}\right)^{1-\xi_{m,\lambda}}H_{1}^{\xi_{m,\lambda}}
= \varpi_{0}H_{1}D_{1}^{\xi_{m,\lambda}-1}\,,
\ee
which, again, is just another expression of (\ref{b8b}). The bounds $1 \leq\lambda < 2$ mean that 
\bel{ximbd}
3/2 \leq \xi_{m,\lambda} < 2\,,
\ee
and so
\bel{H2b}
\shalf\dot{H}_{1} \leq \varpi_{0}\left\{-L^{2} H_{2} + H_{1}D_{1}^{\xi_{m,\lambda}-1}\right\} + \mbox{forcing}\,,
\ee
which, when the $H_{n}$ are adjusted to the $F_{n}$ defined in (\ref{Fndef}) as in \cite{DG02}, becomes 
\bel{H3}
\shalf\dot{F}_{1} \leq \varpi_{0}\left\{-L^{2} F_{2} + F_{1}D_{1}^{\xi_{m,\lambda}-1}\right\} + c_{n}\varpi_{0}Gr\,F_{1}\,.
\ee
Dividing (\ref{H3}) by $F_{1}$ and time averaging, we get \cite{DG02}
\beq{kapest1}
L^{2}\left<\kappa_{2,1}^{2}\right>_{T} \leq \left<D_{1}\right>_{T}^{\xi_{m,\lambda}-1} 
\leq c\,Re^{3\left(\xi_{m,\lambda}-1\right)}\,.
\eeq
Moreover, we can also write 
\bel{kapest2}
\left<\kappa_{2,0}\right>_{T} \leq \left<\kappa_{2,1}\kappa_{1,0}\right>_{T}^{1/2}
\leq \left<\kappa_{2,1}^{2}\right>_{T}^{1/4}\left<\kappa_{1,0}^{2}\right>_{T}^{1/4}\,.
\ee
In \cite{DG02} it was shown that Leray's energy inequality leads to an estimate for $L^{2}\left<\kappa_{1,0}^{2}\right>_{T} 
\leq Re^{1+\varepsilon}$, although from now on we ignore the infinitesimal $\varepsilon >0$. We combine this with (\ref{kapest2}) 
to show that\footnote{\sf To find a good estimate for $\left<\kappa_{n,0}\right>_{T}$ for $n > 2$ using the depletion is a difficult task. 
The estimate for this, found in \cite{DG02} and quoted in (\ref{app8}), is valid in regime II where no depletion result has been used.}
\bel{kapest3a}
\left<\kappa_{2,0}\right>_{T} \leq c\, Re^{\sigma_{m,\lambda}} + O\left(Gr^{1/4}\right)\,,
\ee
where
\bel{kapest3b}
\sigma_{m,\lambda} = \frac{3(\chi_{m,\lambda}-1) + 2(m-1) }{8(m-1)} = (3\lambda + 2)/8\,.
\ee
\par\smallskip%\noindent
To interpret this estimate  physically in terms of statistical 
turbulence theory (restricting attention to forcing at the longest wavelength $\ell = L$), suppose that $Gr$ is high enough and the resulting 
flow is turbulent, ergodic and isotropic enough in the limit $T\to\infty$ that the wave-numbers $\left<\kappa_{n,0}\right>_{T}$ 
may be identified with the moments of the energy spectrum $\mathcal{E}(k)$ according to 
\bel{app1}
\left<\kappa_{n,0}\right>_{T} := \left(\frac{\int_{L^{-1}}^{\infty}k^{2n}\mathcal{E}(k)\,dk}
{\int_{L^{-1}}^{\infty}\mathcal{E}(k)\,dk}\right)^{1/2n}\,.
\ee
The {\em a priori} constraints on $\mathcal{E}(k)$ are that the velocity $U$ and energy dissipation rate $\epsilon$ obey
\bel{app2}
U^{2} =  \int_{L^{-1}}^{\infty}\mathcal{E}(k)\,dk 
\hspace{2cm}
\epsilon =  \int_{L^{-1}}^{\infty}\nu k^{2}\mathcal{E}(k)\,dk\,.
\ee
Suppose also that $\mathcal{E}(k)$ displays an ``inertial range'' in the sense that it scales with a power of $k$ up to an effective 
cut-off wavenumber $k_{c}$. For simplicity, let us write
\bel{app3}
\mathcal{E}(k) = \left\{ 
\begin{array}{cr}
A\,k^{-q}, & \ \ \ L^{-1}\leq k \leq k_{c}\,,\\
0, &       k > k_{c}\,,
\end{array}
\right.
\ee
We also assume that $k_{c}$ diverges as $\nu\to 0$, while $U^{2}$ and $\epsilon$ remain finite, and that $A$ 
depends only upon the energy flux $\epsilon$ and the outer length scale $\ell = L$. Then we have the asymptotic relations 
\bel{app4}
\epsilon \sim \frac{U^3}{L}\hspace{1cm}\mbox{and}\hspace{1cm}
L k_{c} \sim \left(\frac{\epsilon}{\nu^3}\right)^{\frac{1}{9-3q}}
L^{\frac{4}{9-3q}} \sim Re^{\frac{1}{3-q}}\,.
\ee
Then the moments of the spectrum $\left<\kappa_{n,0}\right>_{T}$ satisfy
\bel{app5}
L\left<\kappa_{n,0}\right>_{T}\sim (L\,k_{c})^{1-\frac{q-1}{2n}}
\sim Re^{\frac{1}{3-q} - \frac{1}{2n}\left(\frac{q-1}{3-q}\right)}\,.
\ee
Now let us compare this scaling result with the estimate in (\ref{kapest3a}) for $n=2$ with $q= q_{m,\lambda}$\,: 
this correspondence tells us that
\bel{app6}
q_{m,\lambda} = \frac{12\sigma_{m,\lambda} -5}{4\sigma_{m,\lambda}-1}\,,\qquad\mbox{with}\qquad
q_{m,\lambda} = 3 - \frac{4}{3\lambda}\,.
\ee
In fact, for regime I, $q_{m,\lambda}$ lies between 
\bel{qran1}
5/3 \leq q_{m,\lambda} < 7/3\,.
\ee
The 5/3 at the lower end is the conventional Kolmogorov result which rises to just under 7/3. The cut-off of the 
inertial range as (\ref{app3}) is given by
\bel{qran2}
Lk_{c} \sim Re^{1/(3-q)}\qquad\mbox{so}\qquad Lk_{c,\lambda} \sim Re^{3\lambda/4}\,.
\ee
A resolution length is inbuilt into this formalism\,: the estimate for $L\left<\kappa_{2,0}\right>_{T}$, with an exponent of 
$\sigma_{m}$, can be interpreted as an average length scale. Thus, the first $L\left<\kappa_{1,0}\right>_{T}$ is followed by 
an estimate for $L\left<\kappa_{2,0}\right>_{T}$ at $\chi_{m,\lambda} = m\lambda +1-\lambda$\,: 
\bel{app7}
L\left<\kappa_{1,0}\right>_{T} \leq Re^{1/2}\,,\qquad\qquad
L\left<\kappa_{2,0,\lambda}\right>_{T} \leq Re^{\sigma_{m,\lambda}}\,,
\ee
where $\sigma_{m,\lambda}$ is defined as in (\ref{kapest3b}). This is roughly consistent with scaling arguments 
found in other parts of the literature \cite{YS2005,DYS2008}. 
\par\smallskip%\noindent
In regime II we are forced to revert to the weak solution results in \cite{DG02} where it was shown that for $n \geq 2$,
\bel{app8}
\left<\kappa_{n,0}\right>_{T} \leq c\, Re^{3 - \frac{5}{2n}}\,.
\ee
For $n=2$ this means $\sigma_{m,\lambda} = 7/4$ and thus $q_{m,\lambda}=8/3$. Table 3 summarizes the results for 
both regimes I and II. 
%%%%%%%%%%%%%%%%%%%%%%%%%%%%%
\begin{table}
\bc
\begin{tabular}{||c||c|c|c||}\hline
%Regime I & & & & Regime II\\\hline
$\lambda$ & $1$ &$2$ & $4 $\\\hline
$\sigma_{m,\lambda}=(3\lambda +2)/8$ & $5/8$ & $1$ & 7/4\\\hline
$q_{m,\lambda}=3-4/3\lambda$ & $5/3$ & 7/3 & 8/3\\\hline
$Lk_{c} \leq Re^{3\lambda/4}$ & $Re^{3/4}$ & $Re^{3/2}$ & $Re^{3}$\\\hline
\end{tabular}
\ec
\vspace{-5mm}
\caption{\sf\scriptsize The entries in the second and third columns are the lower and upper bounds of $\sigma_{m,\lambda},~q_{m,\lambda}$ 
and $Lk_{c}$ corresponding to the two concave curves in Fig. 1. The fourth column lists values of these at $\lambda =4$, which is near the 
extreme end of regime II.}
\end{table}
%%%%%%%%%%%%%%%%%%%%%%%%%%%%%
\par\smallskip%\noindent
Interestingly, Sulem and Frisch \cite{SF75} showed that a $k^{-8/3}$ energy spectrum is the borderline steepness 
capable of sustaining an energy cascade. This spectrum corresponds to the extreme limit, where the energy dissipation 
is concentrated on sets of dimension zero (points) in space \cite{Man75,FSN78}. It provides some physical setting
in which to interpret the result of Caffarelli, Kohn \& Nirenberg that the space-time dimension of the Navier-Stokes 
singular set is unity \cite{CKN82}.

%%%%%%%%%%%%%%%%%%%%%%%%%%%%%%%%
\section{\sf\large\textbf{Conclusion}}\label{con}

Three regimes have been identified based on the size of the $D_{m}$ for $m\geq 2$ relative to that of $D_{1}$. Regime I has 
been shown to have a sufficiently depleted nonlinearity that an absorbing ball exists for $D_{1}$. The consequence of this is 
that a global attractor exists, provided solutions remain in regime I.  A diagrammatic description of the relation between the 
three regimes is given below\,: 
%%%%%%%%%%%%
\beq{pic1}\nonumber
\underbrace{........~\mbox{Regime~I}~.......}_{D_{m}\leq D_{1}^{A_{m,\lambda}} - \mbox{\small regular}}~D_{1}^{A_{m,\lambda}}~
\underbrace{...............~\mbox{Regime~II}~.................}_{D_{1}^{A_{m,\lambda}} \leq D_{m} \leq C_{m} D_{1} - \mbox{\small weak solutions}}~
C_{m}D_{1}~\underbrace{...........~\mbox{Regime~III}~..........}_{C_{m}D_{1} < D_{m} - \mbox{\small regular}}
\eeq
Fig. 1 in \S\ref{introduction} depicts these regions in the $D_{m}-D_{1}$ plane. Specifically, the region between $\lambda=1$  and 
$\lambda = 2$ is the region where solutions are regular. The dotted curves within this show the approximate region (not the exact 
trajectories) where the computations of \S\ref{num} lie. 
\par\smallskip%\noindent
These results also prompt the following set of questions.  
\par\smallskip%\noindent
The first question is why should the Navier-Stokes equations choose to operate in regime I, as observed? While the numerical experiments 
in \cite{DGGKPV13} have shown no evidence of a transition from regime I to II, nevertheless, such a transition cannot be discounted for 
different sets of initial conditions or higher Reynolds numbers. This raises the question whether solutions with initial conditions lying in 
regime I remain there for all time?  
If a transition does occur, how might it come about? Are regimes II and III physical in the sense that while mathematically allowable, 
do they represent recognizable turbulent states? Regimes I and II appear to be consistent with the two branches discovered by Lu and Doering 
\cite{LD08} in their maximization of the rate of enstrophy production. The $\xi_{m,\lambda}=1+\shalf\lambda$ result in regime I takes the 
value of 1.75 at $\lambda =1.5$\,: the value at the lower branch in \cite{LD08} is $\xi_{m,\lambda} = 1.78$. The value $\xi_{m,4} =3$ for 
regime 2 is also consistent with $\xi_{m,\lambda}=2.997$ on the upper branch in \cite{LD08}. In a further paper, Schumacher, Eckhardt and 
Doering \cite{SED10} found numerically that $\xi_{m,\lambda}=3/2$. This, however, was derived from an analysis of local concentrations of 
vorticity, not the full volume calculations in this paper. Nevertheless, it is worth pointing out that bounds on $\xi_{m,\lambda}$ are 
\bel{ximlam}
3/2 \leq \xi_{m,\lambda} < 2\,,\qquad\qquad 1\leq \lambda < 2\,,
\ee
so the result in \cite{SED10} lies exactly at the extreme lower bound where the energy spectrum is $q_{m,1}=5/3$. 
\par\smallskip%\noindent
Secondly, what of initial conditions that are the reverse of the observed ordering\,: that is, initial conditions that are in an ascending scale 
and thus satisfy $D_{m} < D_{m+1}$?  A recent numerical experiment by Kerr \cite{RMK2013b} on the $3D$ Euler equations found that, 
in the late stage, the $D_{m}$ did indeed reverse in order to this ascending scale $D_{m} < D_{m+1}$. Then in a further experiment Kerr 
\cite{Kerr2014priv} took this reversed state as initial conditions for the Navier-Stokes equations to discover that the ordering immediately 
switched back again to the descending scale $D_{m+1} < D_{m}$. 
\par\smallskip%\noindent
Thirdly, the magnitude of the vortex stretching term is locally dependent on the angle between $\bom$ and eigenvectors of the strain 
matrix. Overall, this is averaged within the norms buried within the $D_{m}$. Is it possible that a more direct connection could be made in 
the analysis between these results and the work of Constantin and Fefferman and others on the direction of vorticity [55-61,41]? %\cite{CF93,ConstSIAM,Veiga1,Vass09,Ohk09,Veiga2,BosRub13, HouLi06}? 
\par\smallskip%\noindent 
Fourthly, there is a growing body of work on so-called Navier-Stokes-$\alpha$ models, which includes the Leray-$\alpha$, LANS--$\alpha$, 
Clark-$\alpha$ and Bardina models [62-67], %\cite{NSalpha1,NSalpha2,Bard06,Clark05,NSalpha3,NSalpha4}. 
plus the Navier-Stokes-Voight model \cite{Voight08}. All of these models have better regularity properties, 
in differing degrees, than the original Navier-Stokes equations themselves. A comparison between these and the results of regime I might 
be a useful exercise. 
\par\smallskip%\noindent
Finally, the depletion of nonlinearity in regime I is sufficiently strong to suggest that vorticity may be accumulating on low-dimensional sets. 
A generation of graphics has suggested that this is indeed the case\,: vortex sheets rolling up into tubes is typically the situation as a turbulent 
Navier-Stokes flow matures beyond intermediate times. An analytical proof of this poses formidable technical problems as no proof exists for 
the Divergence theorem nor the Sobolev inequalities on a fractal domain 
with evolving fractal boundary conditions.  Given these hurdles all that can be done at present is to re-estimate \textit{formally} 
(\ref{D1est1a}) in $d$-dimensions using dimensional analysis. This suggests that the formal equivalent of (\ref{D1est1a}) is
\bel{D1est2}
\shalf \varpi_{0}^{-1}\dot{D}_{1} \leq - \left(\frac{4-d}{4}\right)\frac{D_{1}^{2}}{E} + c_{d}D_{1}^{\frac{6-d}{4-d}}\,.
\ee
Note that when $d=3$ this reduces to (\ref{D1est1a}). With $\xi_{m,\lambda} = \frac{6-d}{4-d}$ it is easy to calculate the value of $d$ 
corresponding to $\xi_{m,\lambda}=1+ \shalf\lambda$  which is
\bel{dlam}
d_{\lambda} =4\left(1 - \lambda^{-1}\right)\,.
\ee
For instance, this takes the value of  $d_{1.15} \approx 0.52$ when $\lambda = 1.15$ to $d_{4/3} = 1$ when $\lambda = 4/3$. 
This suggests that the low-dimensional set corresponding to a nonlinearity of $D_{1}^{\xi_{m,\lambda}}$ is one which may 
run from being a set of points to tube-like vortical structures [26-32,43]. 
%\cite{VM94,KT05,IGK2009,DYS2008,DY2010,YDS2012,RMK2012a,RMK2013a}. 

%%%%%%%%%%%%%%%
\par\vspace{3mm}\noindent
\textbf{Acknowledgments\,:} DAD acknowledges the computing resources provided by the NSF-supported XSEDE and DOE INCITE 
programs under whose auspices some of these calculations were performed.  DV acknowledges the support of the F\'ed\'eration 
Wolfgang Doeblin and, together, RP and DV acknowledge the support of the ``Indo-French Center for Applied Mathematics'', UMI 
IFCAM -- Bangalore\,; AG and RP thank DST, CSIR and UGC (India) and the SERC (IISC) for computational resources. AG also thanks 
the European Research Council for support under the EU's 7th Framework Programme (FP7/2007-2013)/ERC grant agreement number 
297004.

%%%%%%%%%%%%%%%

\end{document}